\documentclass{aa}  

\usepackage{wasysym}
\usepackage{graphicx}
\usepackage{txfonts}
\usepackage{natbib}
 
\usepackage{hyperref}
\usepackage{textcomp}
\usepackage[flushleft]{threeparttable}

\newcommand{\vel}{Vela SNR}
\newcommand{\XMM}{\textit{XMM-Newton} }

\begin{document} 


\title{X-ray emitting structures in the Vela SNR: Ejecta anisotropies and progenitor stellar wind residuals}

\author{V. Sapienza\inst{1,2}
          \and
           M. Miceli\inst{1,2}
          \and
           G. Peres \inst{1,2}
          \and
          F. Bocchino\inst{2}
          \and
          S. Orlando\inst{2}
          \and
          E. Greco\inst{1,2}
          \and 
          J. A. Combi\inst{3,4}
          \and
          F. Garc\'{i}a\inst{5}
          \and
          M. Sasaki\inst{6}
          }

\institute{Dipartimento di Fisica e Chimica E. Segr\`e, Universit\`a degli Studi di Palermo, Piazza del Parlamento 1, 90134, Palermo, Italy
        \and
        INAF-Osservatorio Astronomico di Palermo, Piazza del Parlamento 1, 90134, Palermo, Italy
        \and
        Instituto Argentino de Radioastronomía (CCT-La Plata, CONICET; CICPBA; UNLP), C.C. No. 5, 1894 Villa Elisa, Argentina
        \and
        Facultad de Ciencias Astronómicas y Geofísicas, Universidad Nacional de La Plata, Paseo del Bosque s/n, 1900 La Plata, Argentina
        \and
        Kapteyn Astronomical Institute, University of Groningen, PO BOX 800, NL-9700 AV Groningen, the Netherlands
        \and
        Dr. Karl Remeis Observatory and ECAP, Universität Erlangen-Nürnberg, Sternwartstr. 7, 96049 Bamberg, Germany
        }
        
\date{}

 
\abstract
{The Vela supernova remnant (SNR) shows several ejecta fragments (or shrapnel) protruding beyond the forward shock, which are most likely relics of anisotropies that developed during the supernova (SN) explosion. Recent studies have revealed high Si abundance in two shrapnel (shrapnel A and G), located in opposite directions with respect to the SNR center. This suggests the possible existence of a Si-rich jet-counterjet structure, similar to that observed in the SNR Cassiopea A.} 
{We analyzed an \XMM observation of a bright clump, behind shrapnel G, which lies along the direction connecting shrapnel A and G. The aim is to study the physical and chemical properties of this clump to ascertain whether it is part of this putative jet-like structure.}
{We produced background-corrected and adaptively-smoothed count-rate images and median photon energy maps, and performed a spatially resolved spectral analysis.}
{We identified two structures with different physical properties. The first one is remarkably elongated along the direction connecting shrapnel A and G. Its X-ray spectrum is much softer than that of the other two shrapnel, to the point of  hindering the determination of the Si abundance; however, its physical and chemical properties are consistent with those of shrapnel A and shrapnel G. The second structure, running along the southeast-northwest direction, has a higher temperature and appears similar to a thin filament. By analyzing the ROSAT data, we have found that this filament is part of a very large and coherent structure that we identified in the western rim of the shell.}
{We obtained a thorough description of the collimated, jet-like tail of shrapnel G in Vela SNR. In addition we discovered a coherent and very extended feature roughly perpendicular to the jet-like structure that we interpret as a signature of an earlier interaction of the remnant with the stellar wind of its progenitor star. The peculiar Ne/O ratio we found in the wind residual may be suggestive of a Wolf-Rayet progenitor for Vela SNR, though further analysis is required to address this point.
}


\maketitle
   

\section{Introduction}
\label{sect1}
Core collapse supernova remnants (SNRs) show complex morphologies that result from intrinsic asymmetries in the supernova (SN) explosion and from the propagation of the explosion shock-waves in very inhomogeneous environments, such as pre-existing stellar winds and molecular clouds.
Therefore, it is difficult to distinguish the role played by the interstellar medium (ISM) inhomogenities from that played by pristine anisotropies in the ejecta in shaping the remnant morphology.
X-ray observations of SNRs are useful diagnostic tools to trace the distribution of the physical and chemical properties of the emitting ejecta and a starting point to reconstruct the details of the explosion mechanism and the structure of the ambient environment surrounding the exploded star.

\vel, the relic from the explosion of a massive star that occurred $\sim11$ kyr ago (\citealt{1993ApJS...88..529T}), is an interesting target to study especially because of its proximity: With a distance of only 280 pc  (\citealt{2003ApJ...596.1137D}), it is the nearest SNR. This makes it possible to resolve the X-ray emission of small-scale structures  spatially and to identify the ejecta to study their properties.

 \begin{figure}
     \centering
         \includegraphics[width=\columnwidth]{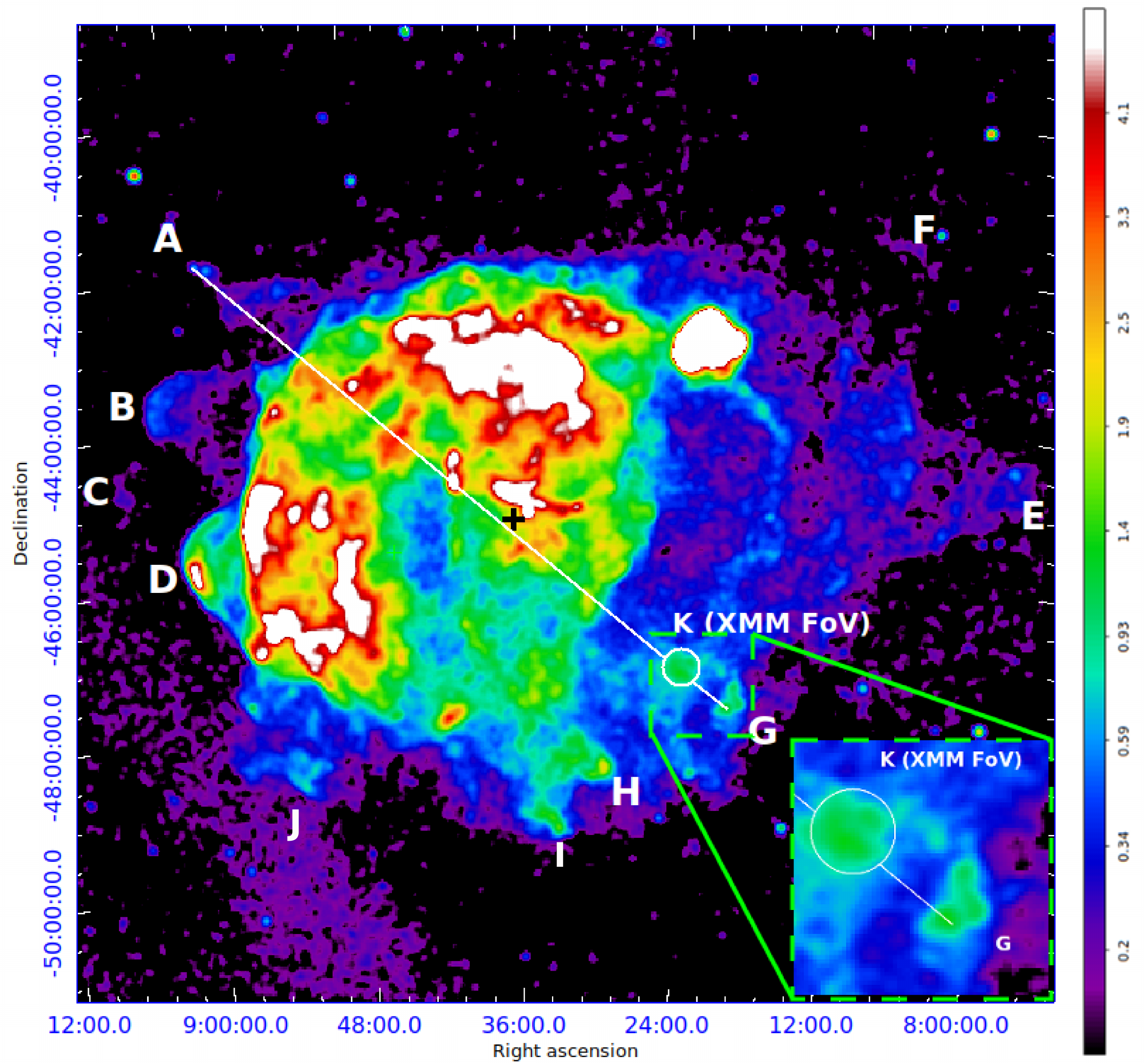}
         \caption{\emph{ROSAT} All Sky Survey (RASS) count image of \vel\ in the $0.1-2.4$ keV energy band in squared root scale.
         The circle K is the region analyzed in this work, marking the \XMM field of view.
         The black cross indicates the position of the Vela pulsar at the explosion time by considering an age of 11000 yr and taking into account the pulsar proper motion estimated by \citet{2001ApJ...561..930C}. We assume it as the explosion center.
         The inset is a close-up view of the same image. 
         } 
         \label{fig:VELA}       
 \end{figure}
 
\citet{1995NYASA.759..196A} identified six X-ray emitting bow-shaped ejecta fragments in regions beyond the forward shock called \textquotedblleft shrapnel,\textquotedblright named from A to F (see Fig. \ref{fig:VELA}).  
X-ray emitting ejecta have also been observed  inside (in projection) the shell (\citealt{2008ApJ...676.1064M}, \citealt{2008ApJ...689L.121L}, \citealt{2018ApJ...865...86S}).
Shrapnel A, which is one of the most distant ejecta from the explosion center, exhibits a pattern of heavy element abundances that are different from that observed in all the other ejecta.
\citet{2006ApJ...642..917K} determined the abundances of the shrapnel by finding a nearly solar abundance for Ne, O, Mg and Fe, and overabundant Si.
The latter is expected to be produced in deeper layers of the progenitor star compared to lower Z elements such as O, Ne, and Mg. 
Therefore, Si-rich ejecta are not expected to be so distant from the explosion center of the remnant, indicating that an inversion of ejecta layers occurred at some point during the remnant evolution.
\citet{2013MNRAS.430.2864M} show, with dedicated 2-D hydrodynamic simulations, that velocity and density contrast with respect to the surrounding ejecta are necessary to make the Si-rich shrapnel overtake the other shrapnel and the forward shock. More recent 3-D magneto-hydrodynamic simulations confirmed the important role played by explosion asymmetries in determining a spatial inversion of ejecta layers (\citealt{2016ApJ...822...22O}, \citealt{2020A&A...642A..67T} and \citealt{2021A&A...645A..66O}).

\citet{2017A&A...604L...5G} analyzed an X-ray luminous knot, named shrapnel G (see Fig. \ref{fig:VELA}), in the southwestern region of \vel.
Shrapnel G is located in the anticenter with respect to shrapnel A. 
\citet{2017A&A...604L...5G} found that the chemical composition of shrapnel G is very similar to that of shrapnel A.
This suggests that shrapnel A and G are part of a jet-counterjet-like event  which has throw away the inner layer of the progenitor star and has made them overcome lighter ejecta and the forward shock (somehow similar to that observed in Cassiopea A, \citealt{2004NewAR..48...61V}). 
To confirm that shrapnel A and G are part of a Si-rich jet-like structure, it is necessary to ascertain the nature of the Si emission and to study the region between the two ejecta knots  in
detail. 

The \emph{ROSAT} image (see Fig. \ref{fig:VELA}) of the whole \vel\ shows that shrapnel G is followed by another bright clump, labeled ``knot K."
This knot seems to lay along the line connecting shrapnel A, shrapnel G, and the center of the SNR. 
This suggests that knot K could be physically linked to shrapnel G, which would prove the existence of a coherent Si-rich feature on a large spatial scale.

In this paper, we present the analysis of an \XMM\ observation of knot K to study its physical and chemical properties, and to ascertain whether it is part of a jet-like structure linking shrapnel A and G.
We complement the analysis of the \XMM data with \emph{ROSAT} archive observations from the western part of \vel's\ shell.

The paper is organized as follows: In Sect. \ref{sect2} we present the data and their analysis;  Sect. \ref{sect3} and Sect. \ref{sect4} show the results of \XMM\ and \emph{ROSAT} data analysis, respectively. Finally, we discuss our results in Sect. \ref{sect5}.


\section{Observations and data analysis}
\label{sect2}
We analyze an \XMM European Photon Imaging Camera (EPIC) observation of \vel\ knot K. 
EPIC consists of a set of three X-ray sensing CCD cameras: two MOS detectors (\citealt{2001A&A...365L..27T}) and a pn detector (\citealt{2001A&A...365L..18S}), operating in the $0.3-10$ keV band.
The \XMM observation was performed from October 07, 2019 to October  08, 2019 (Obs. ID 0841510101, PI: M. Miceli), with pointing coordinates $\alpha_{J2000}=$ 08$^h$23$^m$32.49$^s$ and $\delta_{J2000}=$ -47\textdegree\ 11' 55.8", medium filter, in full frame mode. 
The exposure times are 55 ks, 55 ks, and 51 ks for the MOS1, MOS2, and PN cameras, respectively.
We processed the data with the Science Analysis System (SAS) software, version 18.0.0.

EPIC event files are typically contaminated by soft-protons, that is to say mildly relativistic protons that are detected by the CCD cameras. 
We filtered the event lists for soft-proton contamination with the \texttt{espfilt} task, thus obtaining a screened exposure time of approximately 33 ks for the MOS1, 37 ks for the MOS2, and 19 ks for the pn camera  (only $37\%$ of the total exposure).
We then filtered the event lists retaining only events with \texttt{FLAG=0} and \texttt{PATTERN$\leq$4,12} for pn and MOS cameras, respectively.
We adopted a source detection procedure, using the \texttt{edetect\_chain} SAS task to remove events in circular regions (with radius 18'') around point-like sources. 

All images were background subtracted by adopting the double subtraction procedure described in \citet{2017A&A...599A..45M} to take instrumental, particle, and X-ray background contamination into account.
For this purpose, we used the Filter Wheel Closed (FWC) and Blank Sky (BS) files available at XMM ESAC webpages\footnote{\url{https://www.cosmos.esa.int/web/xmm-newton/filter-closed}\\\url{http://xmm-tools.cosmos.esa.int/external/xmm_calibration//background/bs_repository/blanksky_all.html}}.
All the images presented here are superpositions of the MOS1, MOS2, and pn images, obtained by using the \texttt{emosaic} SAS procedure. 
We produced vignetting-corrected count-rate maps. It is possible to correct for vignetting by dividing the superposed images by the corresponding superposed exposure maps produced by using the \texttt{eexpmap} SAS command. 
The pn exposure maps were scaled by the ratio of the pn$/$MOS effective areas to make MOS-equivalent superposed count-rate maps.
We then smoothed the resulting count-rate maps adaptively in order to reach a user-defined signal-to-noise ratio by adopting the \texttt{asmooth} SAS task.
We performed spatially resolved spectral analysis for all three EPIC detectors.
For this purpose, we corrected the vignetting effect in the spectra using the \texttt{evigweight} SAS command.
For each spectrum, we produced the redistribution matrix file (RMF, with the \texttt{rmfgen} task) and the ancillary response file (ARF, with the \texttt{arfgen} task) and we binned spectra to obtain at least 25 counts per bin.

We also analyzed the \emph{ROSAT} All Sky Survey (RASS) data of the western region of \vel\ to complement the \XMM data analysis.
The RASS was conducted using a Position-Sensitive Proportional Counter (PSPC) that operated in the 0.1-2.4 keV energy band.
The RASS PSPC observation pointing the western part of \vel\ was performed between October 17, 1990 and November 22, 1990, with pointing coordinate $\alpha_{J2000}=8^h15^m00^s$ and $\delta_{J2000}=$-45\textdegree 00'00'' (Observation ID WG932517P\_N1\_SI01.N1) without any filter in survey mode and a total exposure of 23 ks. 
The total  field of view (FoV) of the observation is a  6\textdegree 30' $\times$  6\textdegree 30' box.

The \emph{ROSAT} archive provides processed event lists, distributed as FITS files, and no further data reduction is necessary for the user.
We produced \emph{ROSAT} count maps and spectra using the XSELECT analysis system tools \texttt{extract image} and \texttt{extract spectrum,} respectively.
For the spectral analysis, we used the \texttt{pspcc\_gain1\_256.rmf} RMF file and \texttt{pspcc\_gain1\_256.rsp} on-axis response matrices file to calculate the ARF file for an off-axis region with the FTOOLS command \texttt{pcarf}.
All these files are stored on an HEASARC ftp server\footnote{\url{https://heasarc.gsfc.nasa.gov/FTP/caldb/data/rosat/pspc/cpf/matrices/}}. 
We binned the spectra energy channels in order to have at least 25 counts per channel.
The spectral analysis of \XMM and \emph{ROSAT} spectra was performed with the HEASOFT software XSPEC version 12.9.1 (\citealt{1996ASPC..101...17A}) with the solar abundances table from \citet{1989GeCoA..53..197A}.


\section{\XMM results}
\label{sect3}
\subsection{Images}

\begin{figure}
     \centering
     \includegraphics[width=\columnwidth]{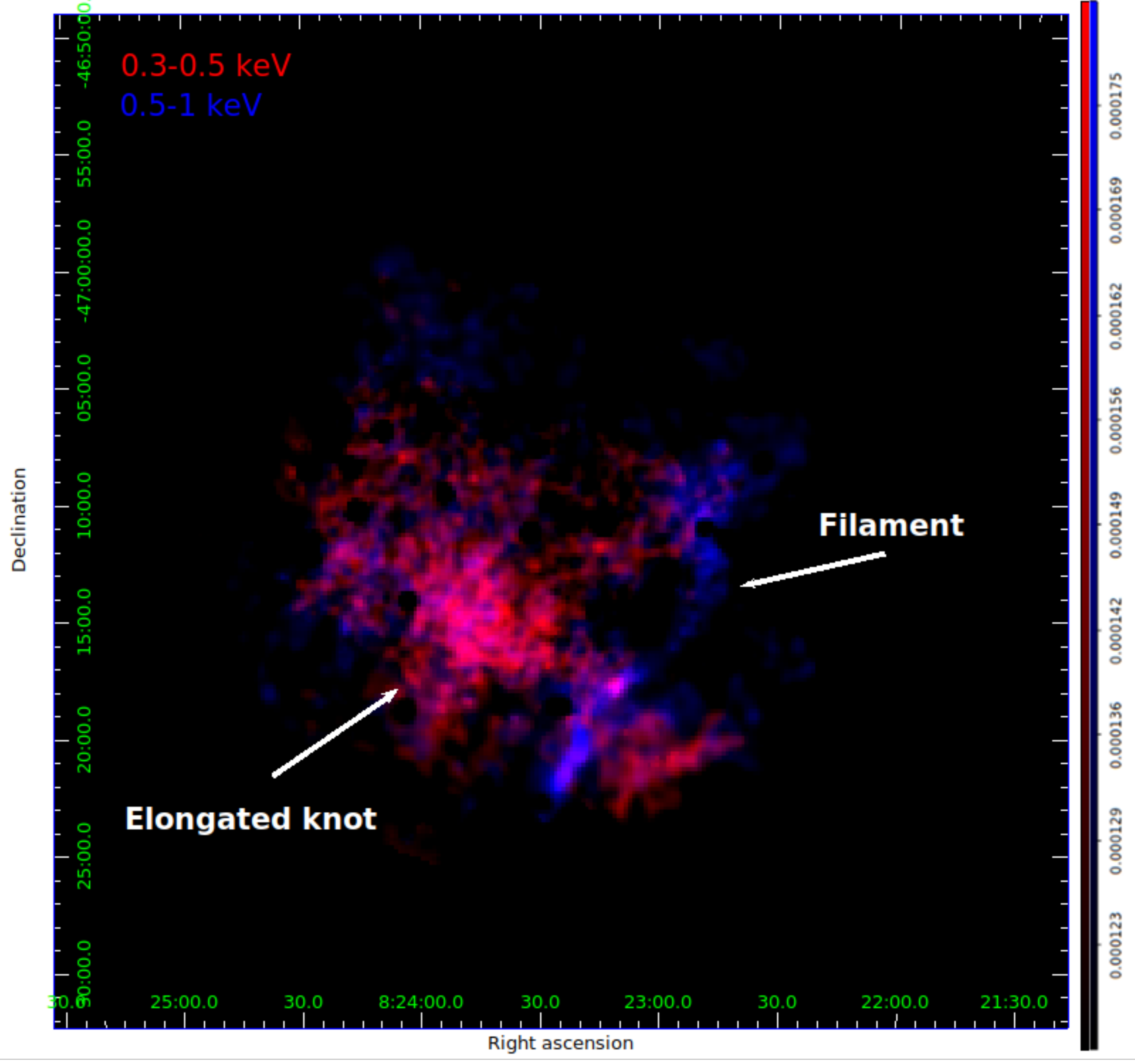}
     \caption{EPIC count-rate color composite image in the $0.3-0.5$ keV band (red) and $0.5-1$ keV band (blue) in linear scale. The bin size is $10''$, and the image was adaptively smoothed to a signal-to-noise ratio of 20.
     North is up and east is to the left.}
     \label{fig:rbimage}
 \end{figure}

In Fig. \ref{fig:rbimage}, we show the EPIC composite count-rate image of the \vel\ knot K, showing the $0.3-0.5$ keV emission in red and the $0.5-1.0$ keV band emission in blue. 
In the $0.3-0.5$ keV band, an elongated knot clearly sticks out. 
This knot seems to be strongly elongated, extending for $\sim20'$ (corresponding to $\sim5\times10^{18}$ cm at a distance of 280 pc) in the northeast-southwest direction, and only $\sim5'$ ( $\sim1\times10^{18}$ cm) on the average in the southeast-northwest direction.
The elongated knot is less visible in the $0.5-1$ keV band, where a narrow filament running in the northwest-southeast direction emerges. 

\begin{figure}
     \centering
     \includegraphics[width=\columnwidth]{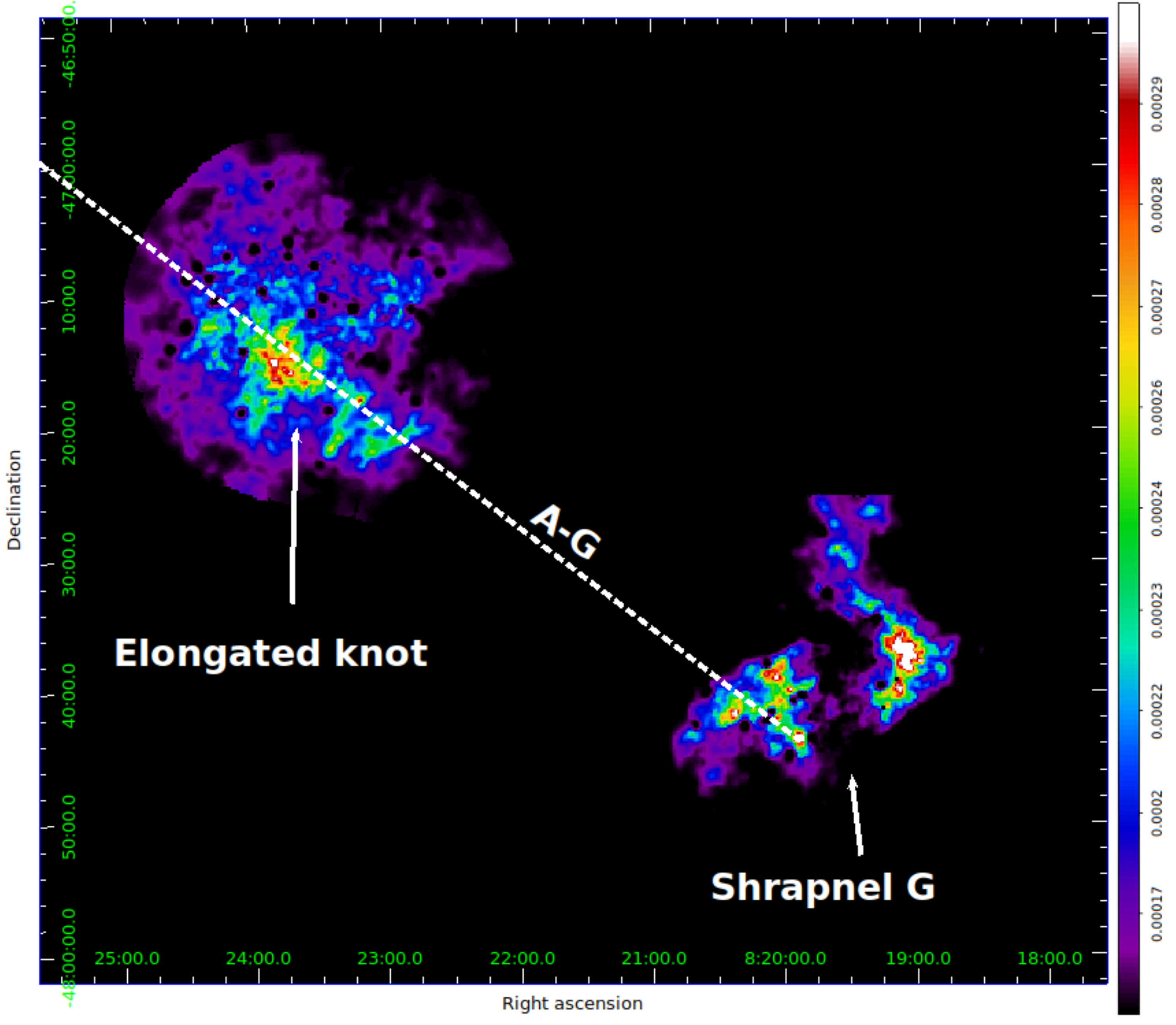}
     \caption{EPIC count-rate image of shrapnel G and  knot K in the $0.3-0.6$ keV energy band in squared scale. 
      The bin size is $10''$ and the image was adaptively smoothed to a signal-to-noise ratio of 20.
     The white dashed line connects shrapnel G and shrapnel A (shrapnel A is outside the field of view of this image, on the opposite side of \vel, see Fig.\ref{fig:VELA}). 
     }
     \label{fig:mosaic}
\end{figure}

In Fig. \ref{fig:mosaic} we show the mosaic of images of the \vel\ shrapnel G and knot K  in the $0.3-0.6$ keV energy band (i.e., the soft energy band adopted by \citealt{2017A&A...604L...5G}) to reveal possible connections between the structures detected in the FoV of knot K and those observed in shrapnel G by \citet{2017A&A...604L...5G}.
The image clearly shows that the elongated knot remarkably lays along the line (indicated by a white dashed line in Fig. \ref{fig:mosaic}, see also Fig. \ref{fig:VELA}) connecting shrapnel A, shrapnel G, and the explosion site.
The explosion site, shown by the black cross in Fig. \ref{fig:VELA}, can be estimated by taking into account the proper motion of the Vela pulsar (\citealt{2001ApJ...561..930C}) and assuming an age of 11000 yr. The excellent alignment between the explosion site, the elongated structure, and the two Si-rich shrapnel is suggestive of a possible physical link between the elongated knot and the Si-rich ejecta. This may indicate that the elongated knot is somehow part of a jet-like structure associated with shrapnel A and G (though its X-ray emission is much softer than that of the two shrapnel).

\subsection{Median photon energy map}

To investigate the thermal distribution of the X-ray emitting plasma, we produced maps of median photon energy ($E_m$) for the pn camera, that is to say maps showing the median energy of the photons detected  pixel-by-pixel in a given energy band. 
These maps not only provide information on the spatial distribution of the X-ray emission spectral hardness of the source, but also on the local value of the equivalent hydrogen column (the higher the absorption, the higher the $E_m$ value). 
The pixels in the map have a size of $20''$ so as to collect more than ten counts per pixel.
We smoothed the map by adopting the procedure described in \cite{2008AdSpR..41..390M}, with $\sigma=60''$.
We verified that the instrumental background does not affect the $E_m$ value, since
it is from 15 up to 30 times lower than the signal. Also, we verified from the FWC file that there are no pieces of evidence of significant fluctuations across the FoV in pn instrumental background.

\begin{figure}
    \centering
    \includegraphics[width=\columnwidth]{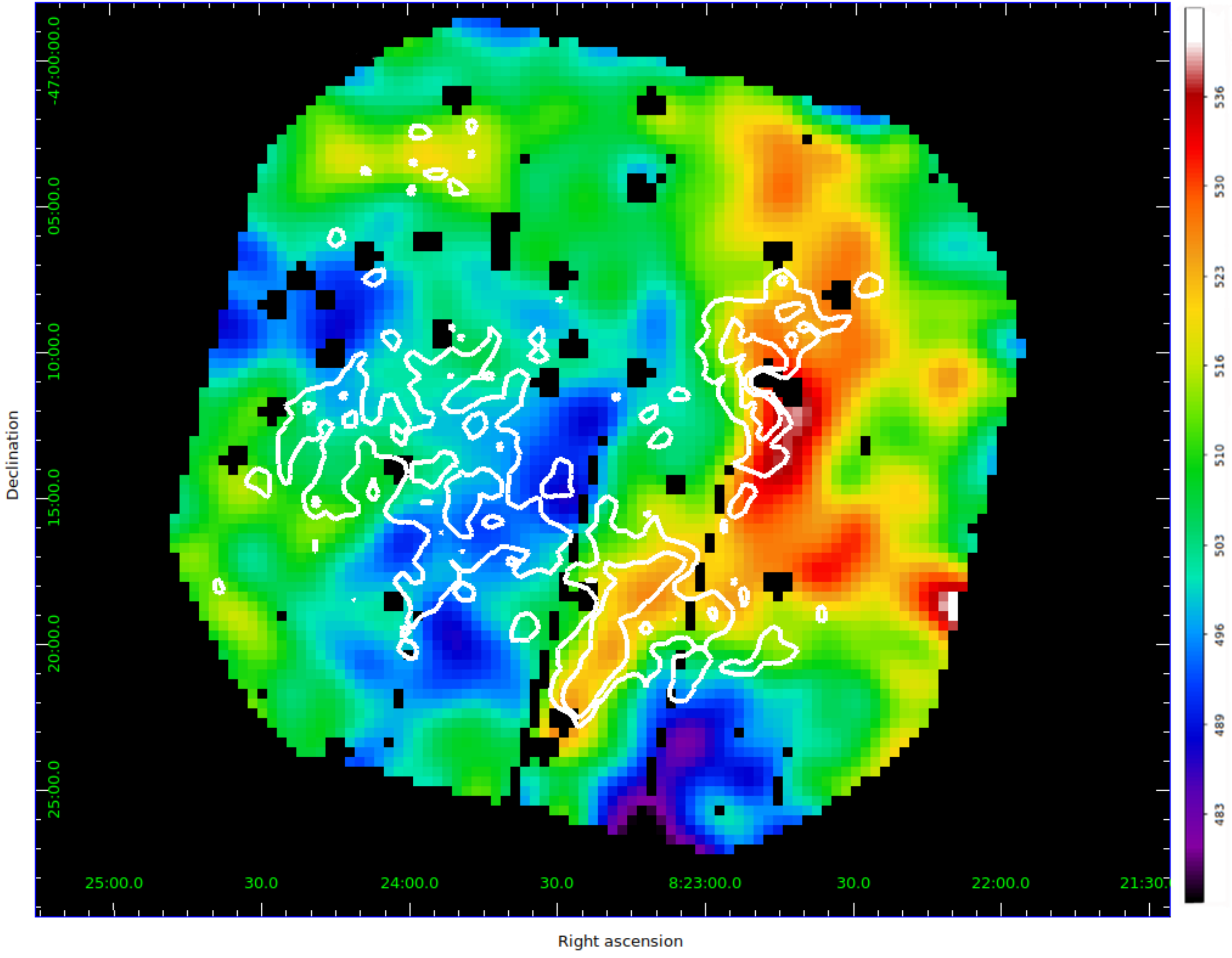}
    \caption{Median photon energy map of the knot K region obtained with EPIC pn data in the $0.3-1.0$ keV energy band in linear scale. The map was smoothed by adopting a Gaussian kernel with $\sigma= 60''$.
    The overlaid white contours indicate the count-rate levels at 50\% and 60\% of the maximum in the $0.5-1.0$ keV energy band.  
    The bin size is $20''$ and the characteristic error for the median energy is $\sim10$ eV.
    }
    \label{fig:mpe03to10}
\end{figure}

In Fig. \ref{fig:mpe03to10} we show the pn $E_m$ map produced in the $0.3-1.0$ keV energy band.
The map clearly shows that the $E_m$ of the elongated knot is lower than that of the filament. Since the interstellar absorption in this region of the Vela SNR is quite uniform (\citealt{2000A&A...362.1083L}), the most natural explanation is that the plasma temperature is higher in the filament than in the elongated knot.
This confirms that the two structures have different physical properties.

\subsection{Spectra}
\label{spec}
 \begin{figure}
      \centering
      \includegraphics[width=\columnwidth]{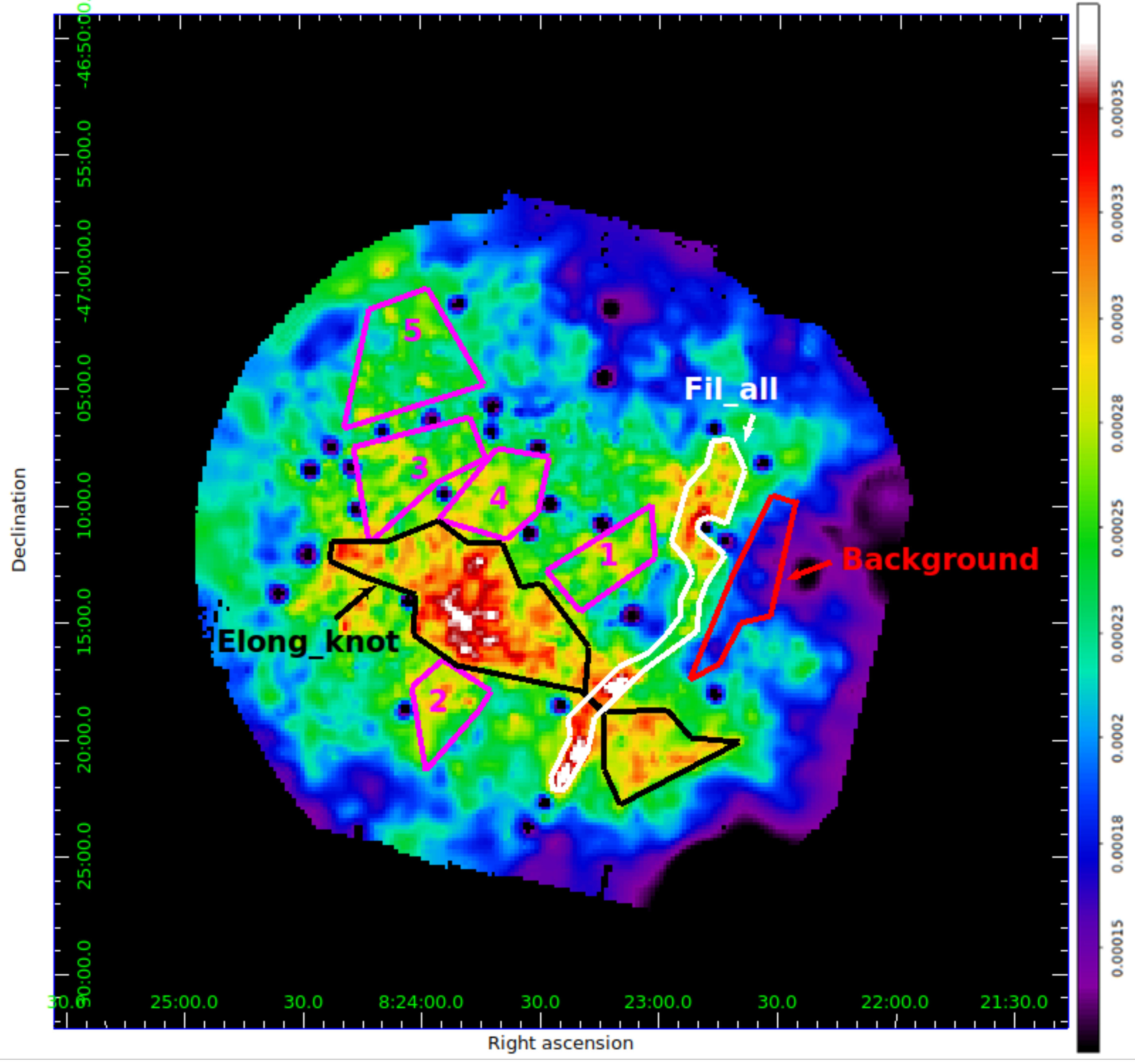}
      \caption{EPIC count-rate image in the $0.3-1.0$ keV energy band in linear scale.
      The bin size is $10''$ and the image was adaptively smoothed to a signal-to-noise ratio of 20.
      Regions selected to extract the spectra of the elongated knot (black polygon) and of the filament (white polygon) are superimposed. The region selected for background extraction is shown in red. The regions selected to extract control spectra are shown in magenta (see Sect. \ref{spec}).}
      \label{fig:specreg}
 \end{figure}  
 
The analysis described in the previous subsection allowed us to identify X-ray emitting plasma structures with homogeneous physical properties.
To further investigate the  physical and chemical properties of the plasma, we performed a spatially-resolved spectral analysis.
We first focused  on the elongated knot which appears similar to a trailing wake of shrapnel G (see Fig. \ref{fig:mosaic}).

We selected a polygonal spectral-extraction region labeled \emph{Elong\_knot}, including the whole jet-like structure, indicated by the black polygon in Fig. \ref{fig:specreg}.
We extracted the spectrum of the background from the red region shown in Fig. \ref{fig:specreg}, characterized by a very low surface brightness.
By selecting other background regions (within the blue
areas of Fig \ref{fig:specreg}), we verified that the results of the spectral analysis do not change significantly: The best fit parameters are all consistent within 1$\sigma$.
The spectra of all cameras above 1.3 keV are dominated by the background, also including the Si fluorescence emission lines of the MOS instrumental background.
We verified that the results of the spectral analysis do not change significantly by performing the analysis in the $0.3-2$ keV band or in the $0.3-1.3$ keV band. Thus we performed the analysis in the $0.3-1.3$ keV band to maximize the signal-to-noise ratio. 

\begin{figure}
    \centering
    \includegraphics[width=\columnwidth]{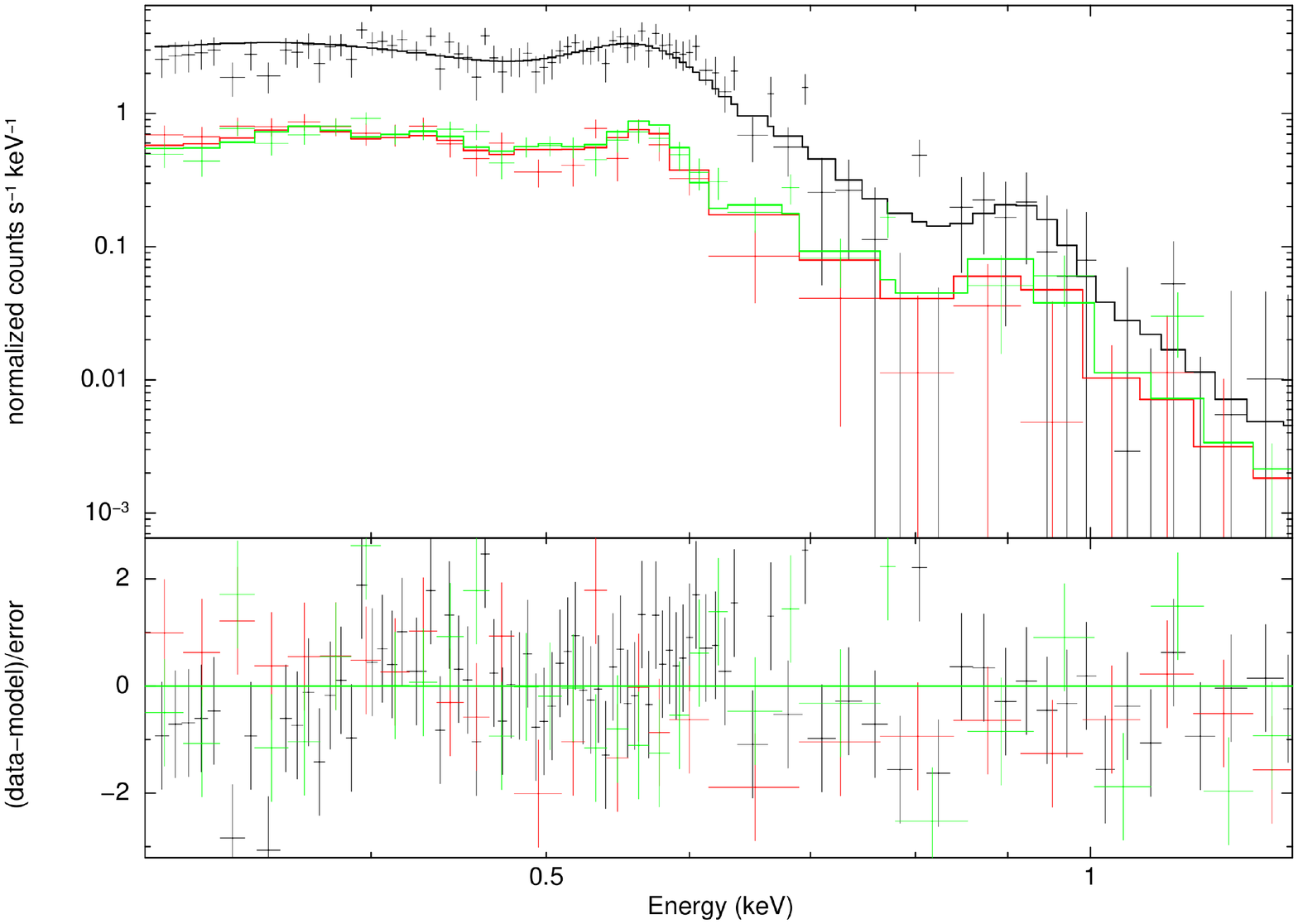}
    \caption{EPIC spectra (pn upper, MOS1,2 lower) extracted from the \emph{Elong\_knot} region (shown in Fig. \ref{fig:specreg})
    with the corresponding CIE best-fit model and residuals in the $0.3-1.3$ keV band.}
    \label{fig:knotspec}
\end{figure}

\begin{table}
\centering
\caption{Best-fit parameters for the region \emph{Elong\_knot} shown in Fig. \ref{fig:specreg} with the CIE best-fit model (\texttt{vapec}). 
}

        \label{tab:knotbest}  
        \medskip

        \begin{tabular} {cc}
                \hline \hline
                Parameter&Elong\_knot
                \\
                \hline
                \vspace{0.1cm}
                $^a N_{\text{H}}$ ($10^{22}$ cm$^{-2}$)&$>0.05$

                \\
                \vspace{0.1cm}
                $kT$ (keV)&$0.124_{-0.005}^{+0.006}$

                \\
                \vspace{0.1cm}
                O/O$_{\astrosun}$&$0.36_{-0.05}^{+0.07}$

                \\
                \vspace{0.1cm}
                Ne/Ne$_{\astrosun}$&1$^b$

                \\
                \vspace{0.1cm}
                $^c n^2l$ ($10^{18}$ cm$^{-5}$)&$1.7_{-0.2}^{+0.1}$

                \\
                \vspace{0.1cm}
                $\chi^2/d.o.f.$ &327.97/300
            \\
                \hline
           \end{tabular}
           
\begin{threeparttable}   
\begin{tablenotes}
  \item Errors are at the 90\% confidence level.
  \item \textbf{Notes.} $^a$Upper limit fixed to 0.06. $^b$Fixed value. $^c$Emission measure per unit area.
   \end{tablenotes}
   \end{threeparttable}
\end{table}

The elongated knot spectrum shows thermal features, namely emission line complexes of O VII at $\sim0.57$ keV and of Ne IX at $\sim0.92$ keV (see Fig. \ref{fig:knotspec}). 
We fit the spectra with a model of isothermal optically-thin plasma in collisional ionization equilibrium (CIE) (\texttt{vapec} model in XSPEC) with nonsolar abundances. We also included the effects of photoelectric absorption by ISM (\texttt{tbabs} model in XSPEC).
By letting the O abundance free to vary, the fit provides $\chi^2/d.o.f.=328.14/300$.
We verified that the quality of the fit does not significantly improve by letting the Ne abundance free to vary.
Moreover, the fitting procedure is not sensitive to the Fe abundance, probably because the plasma temperature  ($T\sim1.5\times10^6$ K) is too low for significant Fe L lines to be emitted in the $0.7-1.3$ keV energy band.
Best-fit values are shown in Tab. \ref{tab:knotbest}. Error bars are at a 90\% confidence level.

\begin{figure}
    \centering
    \includegraphics[width=\columnwidth]{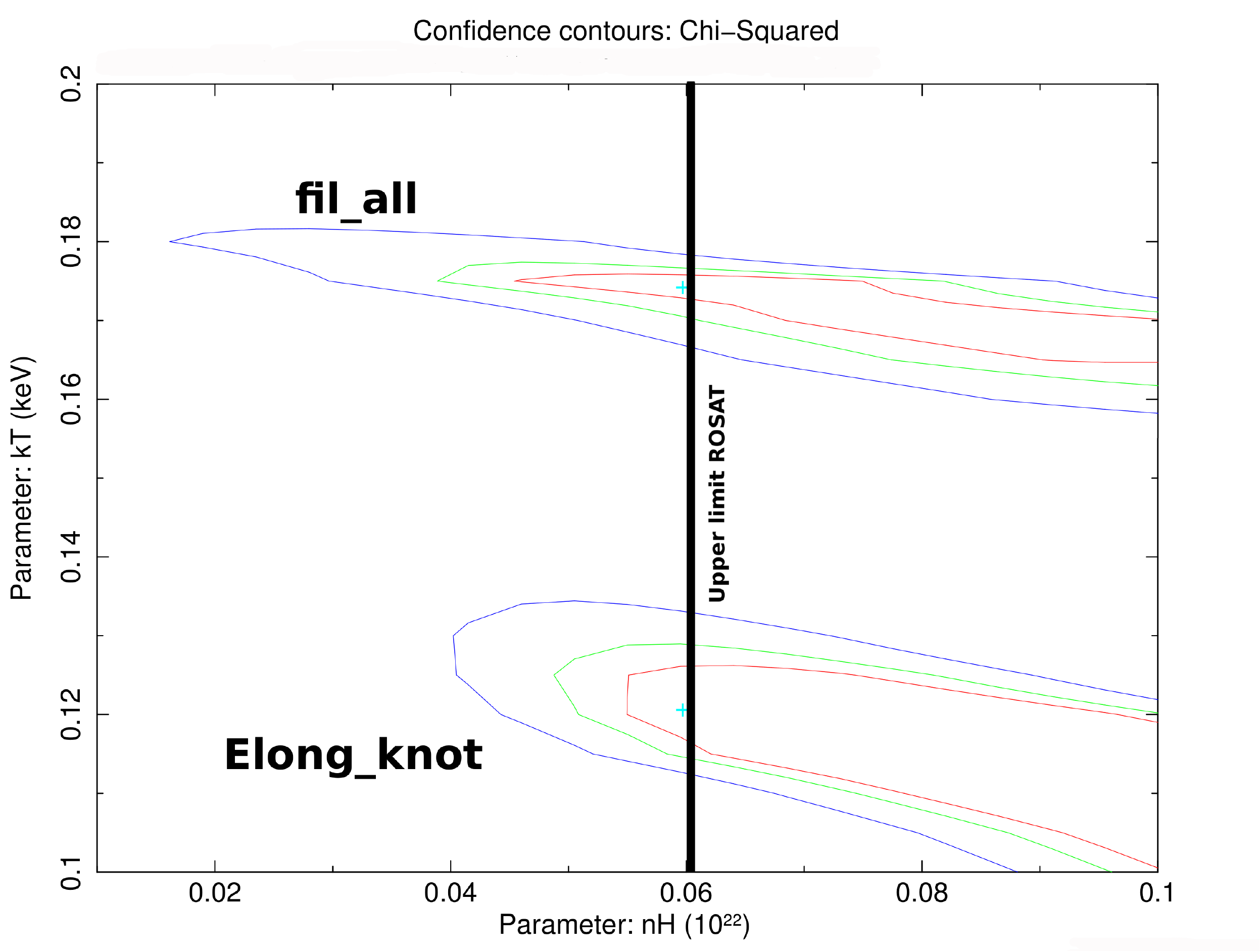}
    \caption{68\% (red), 95\% (green), and 99.5\% (blue) contour levels of the $N_{\text{H}}$ vs. $kT$ best fit values, as derived from the spectral analysis of the \emph{fil\_all} and \emph{Elong\_knot} spectra. The black line shows the upper limit for $N_{\text{H}}$ found by \citet{2000A&A...362.1083L}. }
    \label{fig:nhvskt}
\end{figure}

The best fit temperature is somehow entangled with the best fit value of the column density (as shown in Fig. \ref{fig:nhvskt}), which is forced to be lower than $6\times10^{20}$ cm$^{-2}$, in agreement with the findings by \citet{2000A&A...362.1083L} in this region of the shell.
Though the spectral analysis of shrapnel G makes it possible to constrain the chemical abundances of more elements than that of knot K,  the Ne (and Mg) to O abundance ratio that we found in the elongated knot  (Ne/O = Mg/O $ = 2.6_{-0.4}^{+0.5}$ with Ne and Mg abundances fixed to the solar values) is similar to that found in the ejecta in other regions of the Vela SNR.\ This includes shrapnel G (\citealt{2017A&A...604L...5G}), shrapnel A (\citealt{2006ApJ...642..917K}), shrapnel D (\citealt{2005PASJ...57..621K}), the ejecta found in the RegNe region (\citealt{2008ApJ...676.1064M}), and those in the Vela X region (\citealt{2008ApJ...689L.121L}).
Though an ISM origin cannot be firmly excluded, this result seems to support an ejecta origin for the elongated knot.

 \begin{table}[t!]
        \centering
        \caption{Best-fit parameters for spectra from regions labeled from 1 to 5 shown in Fig. \ref{fig:specreg}.}
        \label{tab:knotpossbest}
        \medskip
    \resizebox{\columnwidth}{!}{
        \begin{tabular}{cccccc}
                \hline \hline
                        Parameter&1&2&3&4&5
                \\
                \hline
                \vspace{0.1cm}
                kT (keV)&$0.108_{-0.012}^{+0.009}$&$0.109_{-0.015}^{+0.005}$&$0.108_{-0.015}^{+0.004}$&$0.109_{-0.009}^{+0.007}$&$0.108_{-0.017}^{+0.003}$
                \\
                \vspace{0.1cm}
                O/O$_{\astrosun}$&$0.40_{-0.1}^{+0.334800}$&$0.66_{-0.16}^{+0.19}$&$0.5_{-0.1}^{+0.2}$&$0.56_{-0.15}^{+0.15}$&$0.6_{-0.1}^{+0.4}$
                \\
                \vspace{0.1cm}
                $n^2l$ ($10^{18}$ cm$^{-5}$) &$1.4_{-0.1}^{+0.5}$&$2.3_{-0.4}^{+0.6}$&$1.5_{-0.2}^{+0.6}$&$1.6_{-0.2}^{+0.3}$&$1.3_{-0.1}^{+0.6}$
                \\              \vspace{0.1cm}
                $\chi^2$/d.o.f. &255.45/225&175.15/168&260.62/252&284.48/239&267.15/243
                \\
                \hline
        \end{tabular}}
\begin{threeparttable}   
\begin{tablenotes}
  \item Errors are at the 90\% confidence level.
  \item 
  \item \textbf{Notes.} $N_{\text{H}}$ value fixed to 0.06 and Ne/Ne$_{\astrosun}$ value is fixed to 1.
   \end{tablenotes}
   \end{threeparttable} 
\end{table}

We also extracted spectra from five regions surrounding the \emph{Elong\_knot}  (labeled reg 1-5 in Fig. \ref{fig:specreg}) and found that the plasma temperature therein is always significantly lower than that of the knot.
Moreover, the Ne/O abundance ratio is systematically (sometimes significantly) lower than that of the elongated knot (see Tab. \ref{tab:knotpossbest}). 
These results clearly show that all these regions do not belong to the elongated knot, which is indeed confined to a narrow, jet-like stripe along the direction connecting shrapnel A and shrapnel G.

\cite{2017A&A...604L...5G} modeled the spectrum of shrapnel G with an absorbed isothermal component in nonequilibrium of ionization (NEI) and nonsolar abundances. We checked if a similar scenario can be adopted for the elongated knot K by fitting its spectrum with the same model as that in \cite{2017A&A...604L...5G} by letting only the plasma temperature and the $n_H$ free to vary. 
We thus obtained a good fit to the \emph{Elong\_knot} spectrum ($\chi^2/d.o.f.=435.55/302$).  
Nevertheless, the quality of the fit is clearly worse than that obtained by the model in CIE.
In any case, we found that the two spectral models adopted (CIE and NEI) clearly show that the plasma temperature of the elongated knot is significantly lower than that of shrapnel G (see Tab. \ref{tab:knotbest} and Tab. \ref{tab:garciabest}), as we expected.
Because of its interaction with the medium swept, the head of a jet-like structure (shrapnel G) is hotter than its tail (knot K), that is interacting with an expanding medium.

At the low temperature of the elongated knot, the emissivity of the Si XIII emission line is extremely low and this hampers the emergence of the Si emission line above the continuum component of the spectrum and above the background. This hinders the possibility of obtaining an accurate measure of the Si abundance in the elongated knot, given that the emerging spectrum is actually insensitive to this parameter.
We have repeated the spectral analysis of the elongated knot either in CIE or in NEI by also including the spectral data points in the $1.3-2$ keV energy band to include the energy bins corresponding to the Si XIII emission line (around 1.8 keV). We let the Si abundance free to vary in the fitting procedure. We found that the Si abundance is almost unconstrained (as expected), obtaining a best fit value of $6\pm4$ in NEI, and the abundance constrained to be $<2$ in CIE. 
In any case, the Si abundance is consistent with that found in shrapnel G by \citet{2017A&A...604L...5G} and in shrapnel A by \citet{2006ApJ...642..917K}. 
Therefore, although it is not possible to prove that the jet-like structure is Si-rich, the spectral analysis shows that a spectrum of Si-rich plasma is perfectly consistent with the observed spectrum and that the jet-like structure could have the same Si abundance as shrapnel A and shrapnel G.

 \begin{table}
        \centering
        \caption{Best-fit parameters for region \emph{Elong\_knot} shown in Fig. \ref{fig:specreg} obtained with the NEI model compared with those found in shrapnel G by \cite{2017A&A...604L...5G}.
        }
        \label{tab:garciabest}  
          \medskip

          \begin{tabular} {ccc}
                \hline \hline
                Parameter &Elong\_knot&Shrapnel G
                \\
                \hline
                \vspace{0.1cm}
                $N_\text{H}$ ($10^{22}$ cm$^{-2}$)&0.06$^\star$&$0.022_{-0.014}^{+0.014}$
                \\
                \vspace{0.1cm}
                $kT$ (keV)&0.113$_{-0.006}^{+0.011}$&$0.49_{-0.04}^{+0.04}$
                \\
                \vspace{0.1cm}
                $\tau$ ($10^{10}$ s cm$^{-3}$)&3.1$^\star$&3.1$_{-0.3}^{+0.3}$
                \\
                \vspace{0.1cm}
                C/C$_{\astrosun} = N/N_{\astrosun} = O/O_{\astrosun}$&0.47$^\star$&$0.47_{-0.10}^{+0.10}$
                \\
                \vspace{0.1cm}
                Ne/Ne$_{\astrosun}$&1.33$^\star$&$1.3_{-0.2}^{+0.2}$
                \\
                \vspace{0.1cm}
                Mg/Mg$_{\astrosun}$&0.92$^\star$&$0.9_{-0.2}^{+0.2}$
                \\
                \vspace{0.1cm}
                Si/Si$_{\astrosun}$&2.24$^\star$&$2.2_{-0.9}^{+0.9}$
                \\
                \vspace{0.1cm}
                Fe/Fe$_{\astrosun}$&0.29$^\star$&$0.29_{-0.08}^{+0.08}$
                \\
                \vspace{0.1cm}
                $n^2l$ ($10^{18}$ cm$^{-5}$)&$1.7_{-0.2}^{+0.1}$&$0.21_{-0.06}^{+0.06}$
                \\
                \vspace{0.1cm}
                $\chi^2$/d.o.f. &435.55/302&635.91/451
            \\
                \hline

          \end{tabular}
          \begin{threeparttable}   
\begin{tablenotes}
\item Errors are at  a 90\% confidence level.
  \item \textbf{Notes.} $^\star$Fixed value.
   \end{tablenotes}
   \end{threeparttable}
\end{table}

\begin{figure}[!t]
    \centering
    \includegraphics[width=\columnwidth]{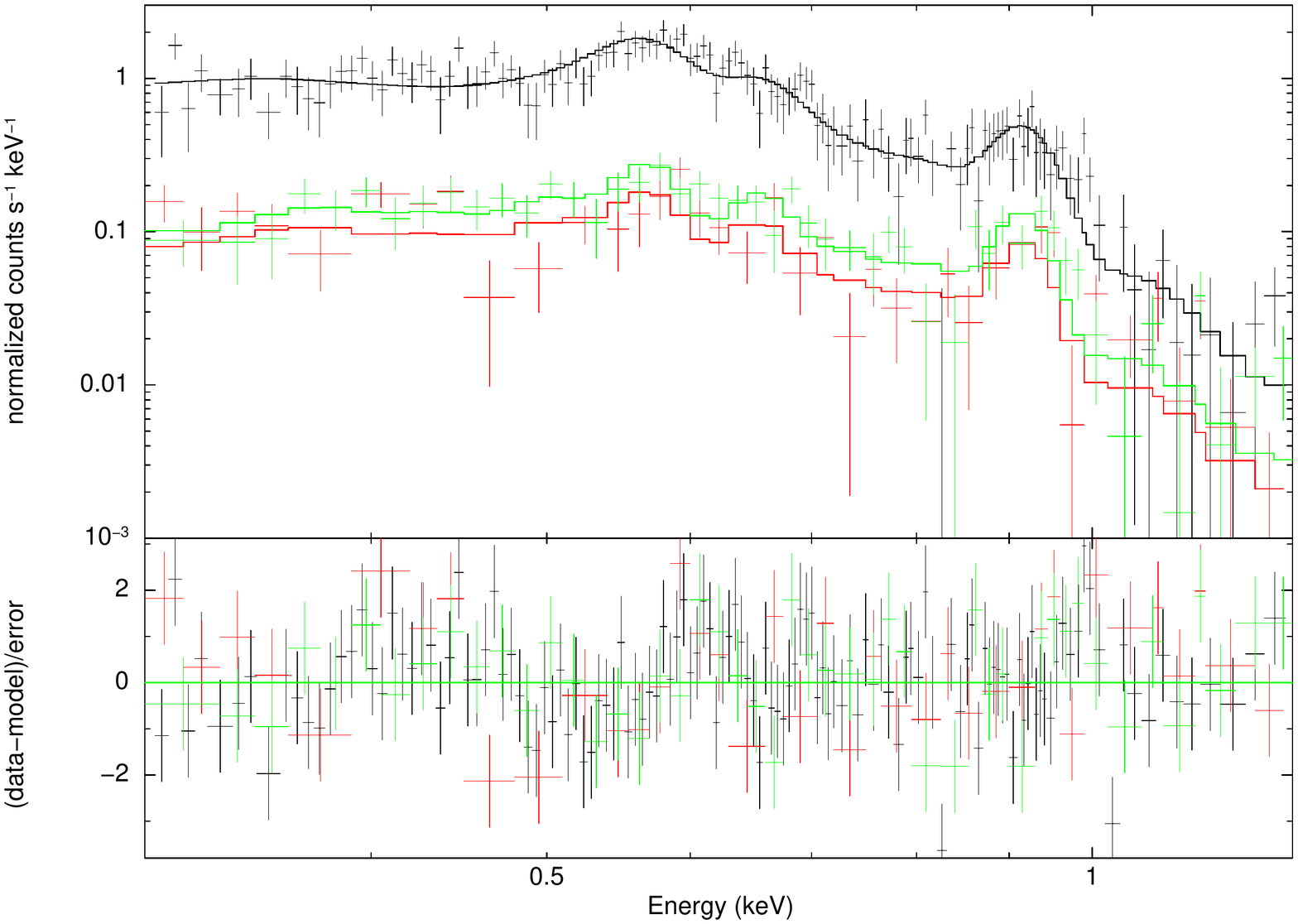}
    \caption{EPIC spectra (pn upper, MOS1,2 lower) extracted from the \emph{fil\_all} region (shown in Fig. \ref{fig:specreg})
    with the corresponding CIE best-fit model and residuals in the $0.3-1.3$ keV band.}
    \label{fig:filspec}
\end{figure}

The second X-ray emitting structure that emerges in Fig. \ref{fig:rbimage} has the shape of a narrow filament.
Therefore, we extracted the spectrum from a large region indicated by the white polygon, labeled \emph{fil\_all}, in Fig. \ref{fig:specreg}, including the whole filament.
We adopted the same background spectrum as that used for the elongated knot.
The spectrum of the region shows thermal emission, characterized by emission lines from O VII at $\sim0.57$ keV, from O VIII at $\sim0.65$ keV, and from Ne IX at $\sim0.92$ keV (see Fig. \ref{fig:filspec}).
To fit the spectrum, we first adopted a model of isothermal plasma in CIE (\texttt{vapec}) with solar abundances, including the effects of photoelectric absorption by the ISM (\texttt{tbabs}).
We fixed the $N_\text{H}$  to the best-fit value found for the elongated knot.
By letting the O and Ne abundance free to vary, the fit provides $\chi^2/d.o.f. = 337.67/276$.
Best-fit values are shown in Tab. \ref{tab:filbest}. Error bars are at a 90\% confidence level.
EPIC spectra with the corresponding best-fit model and residuals are shown in Fig. \ref{fig:filspec}.
The best-fit temperature in the \emph{fil\_all} region is significantly higher than that of the elongated knot (see Fig. \ref{fig:nhvskt}, Tab. \ref{tab:knotbest} and Tab. \ref{tab:filbest}). This confirms that the filament is indeed hotter than the elongated knot, as suggested by the image analysis.
The Ne/O ratio is significantly higher than that found for the knot (Ne/O$_{fil\_all}=4.0^{+0.6}_{-0.8}$ versus Ne/O$_{Elong\_knot}=2.6^{+0.4}_{-0.5}$), and it is also higher than that in the previously mentioned ejecta fragments.
The different temperatures and abundances confirm the different nature of the two plasma structures.

\begin{table}
        \centering
        \caption{Filament spectra best-fit parameters with the CIE best-fit model (\texttt{vapec}).}
        \label{tab:filbest}
        \medskip

        \begin{tabular}{cc}
                \hline \hline
                        Parameter&fil\_all
                \\
                \hline
                \vspace{0.1cm}
                kT (keV)&$0.174_{-0.002}^{+0.002}$
                \\
                \vspace{0.1cm}
                O/O$_{\astrosun}$&$0.48_{-0.06}^{+0.07}$
                \\
                \vspace{0.1cm}
                Ne/Ne$_{\astrosun}$&$1.9_{-0.3}^{+0.4}$
                \\
                \vspace{0.1cm}
                $n^2l$ ($10^{18}$ cm$^{-5}$)&$0.93_{-0.12}^{+0.12}$
                \\
                \vspace{0.1cm}
                $\chi^2$/d.o.f. &337.67/276
                \\
                \hline
        \end{tabular}
                  \begin{threeparttable}   
\begin{tablenotes}
\item Errors are at a  90\% confidence level.
   \end{tablenotes}
   \end{threeparttable}
\end{table}


\section{ROSAT results}
\label{sect4}
\subsection{Images}

The filament that we found in the \emph{XMM-Newton} observation runs from the upper right to the lower left corners of the instrument FoV (see the blue structure in Fig. \ref{fig:rbimage}) and it may therefore  be a part of a large-scale structure extending beyond the \emph{XMM-Newton} FoV.
We thus investigated the nature of this feature by analyzing RASS observations covering the whole western part of \vel.

 \begin{figure}
     \centering
     \includegraphics[width=0.8\columnwidth]{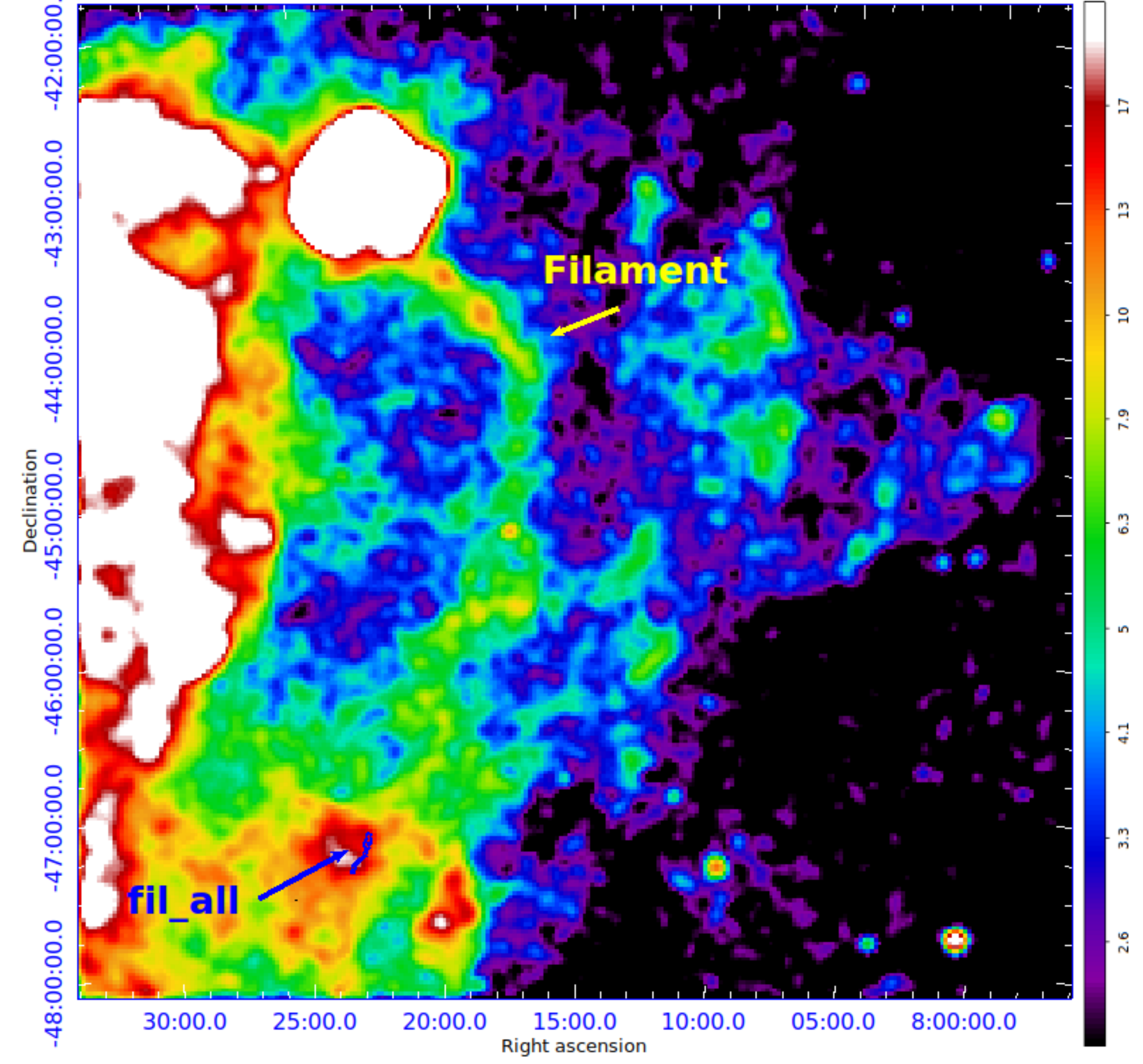}
     \includegraphics[width=0.8\columnwidth]{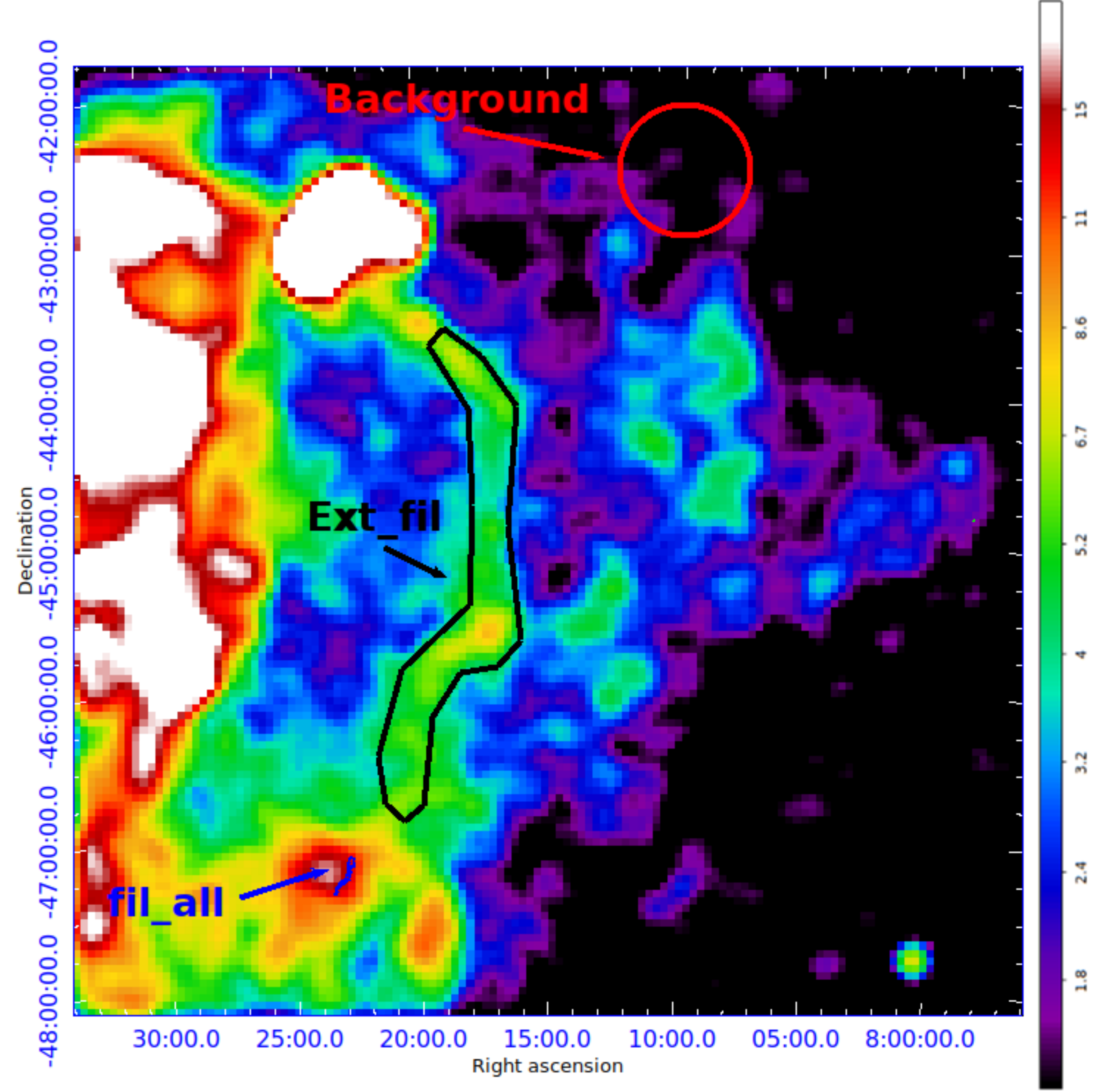}
     \includegraphics[width=0.8\columnwidth]{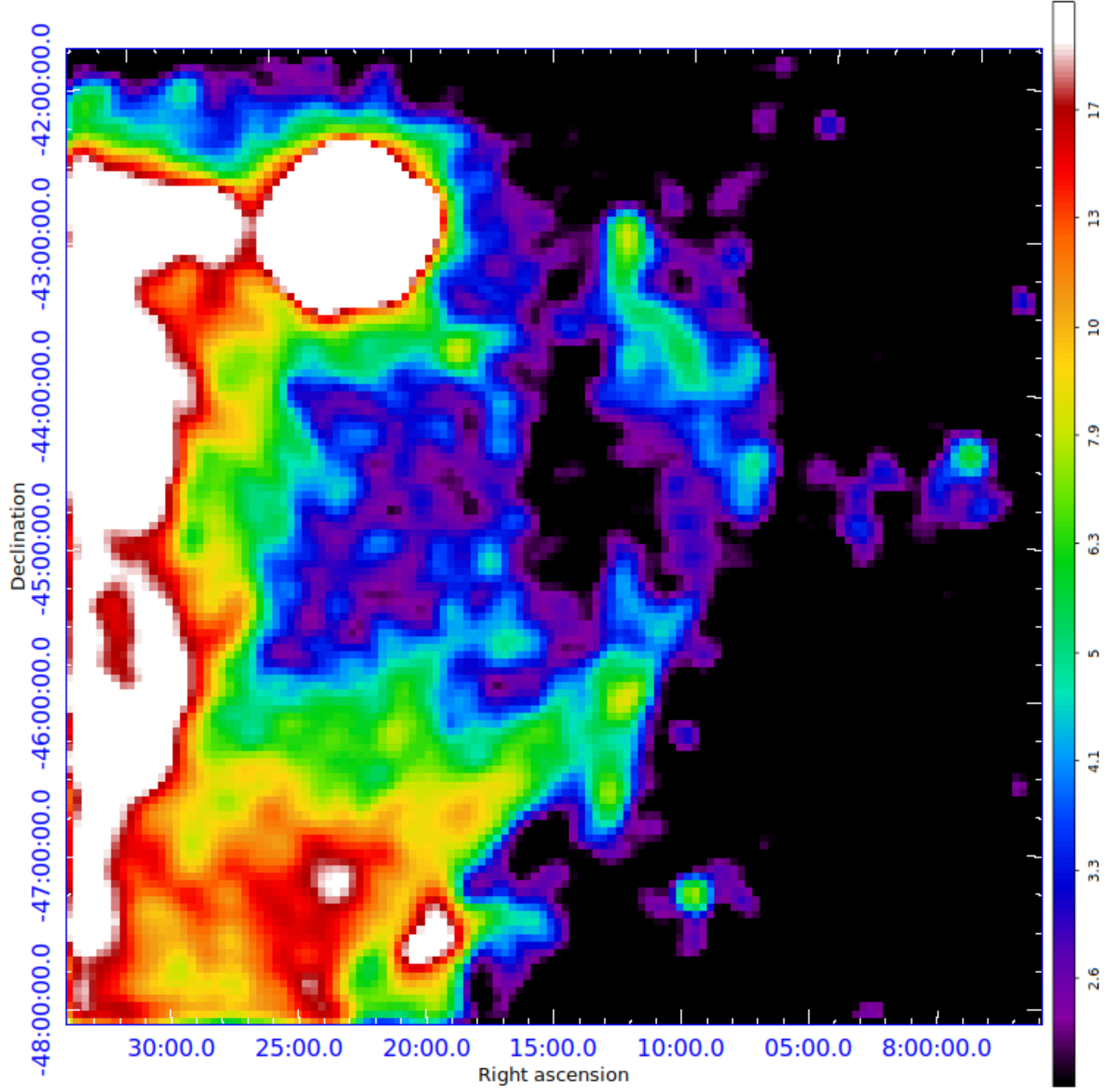}   
     \caption{
        \emph{Top panel}: \emph{ROSAT} PSPC map of photon counts in the $0.1-2.4$ keV energy band in squared root scale. The bin size is $1'$. The map was smoothed through the convolution with a Gaussian with $\sigma=3'$ (3 pixel). The \emph{fil\_all} spectral region (see Fig. \ref{fig:specreg}) is superimposed in blue.
        \emph{Center panel}: \emph{ROSAT} PSPC map of photon counts in the $0.3-0.5$ keV energy band in squared root scale, smoothed with $\sigma=9'$. The bin size is $3'$. The red circle indicates the background spectrum region, and the black polygon is the filament spectrum extraction region.
        The \emph{fil\_all} spectral region is superimposed in blue.
        \emph{Bottom panel}: \emph{ROSAT} PSPC map of photon counts in the $0.5-1.0$ keV energy band in squared root scale, smoothed with $\sigma=9'$. The bin size is $3'$.
        }
     \label{fig:ROSAThardsoft}
\end{figure}

In Fig. \ref{fig:ROSAThardsoft} we show maps of photon counts of the western part of \vel\ in the broad \emph{ROSAT} bandpass (i.e., $0.1-2.4$ keV) and in the $0.3-0.5$ keV (soft) and $0.5-1.0$ keV (hard) energy bands. 
In the \emph{ROSAT} broad and soft band images, an extended filament running from  north (at a position approximately corresponding to the projected location of the Puppis A SNR) to the south, and extending down exactly to the knot K region, is clearly visible.
The extended filament blends with the surrounding emission in the $0.5-1.0$ keV image.

\subsection{Median photon energy maps}

\begin{figure}[!ht]
     \centering
     \includegraphics[width=\columnwidth]{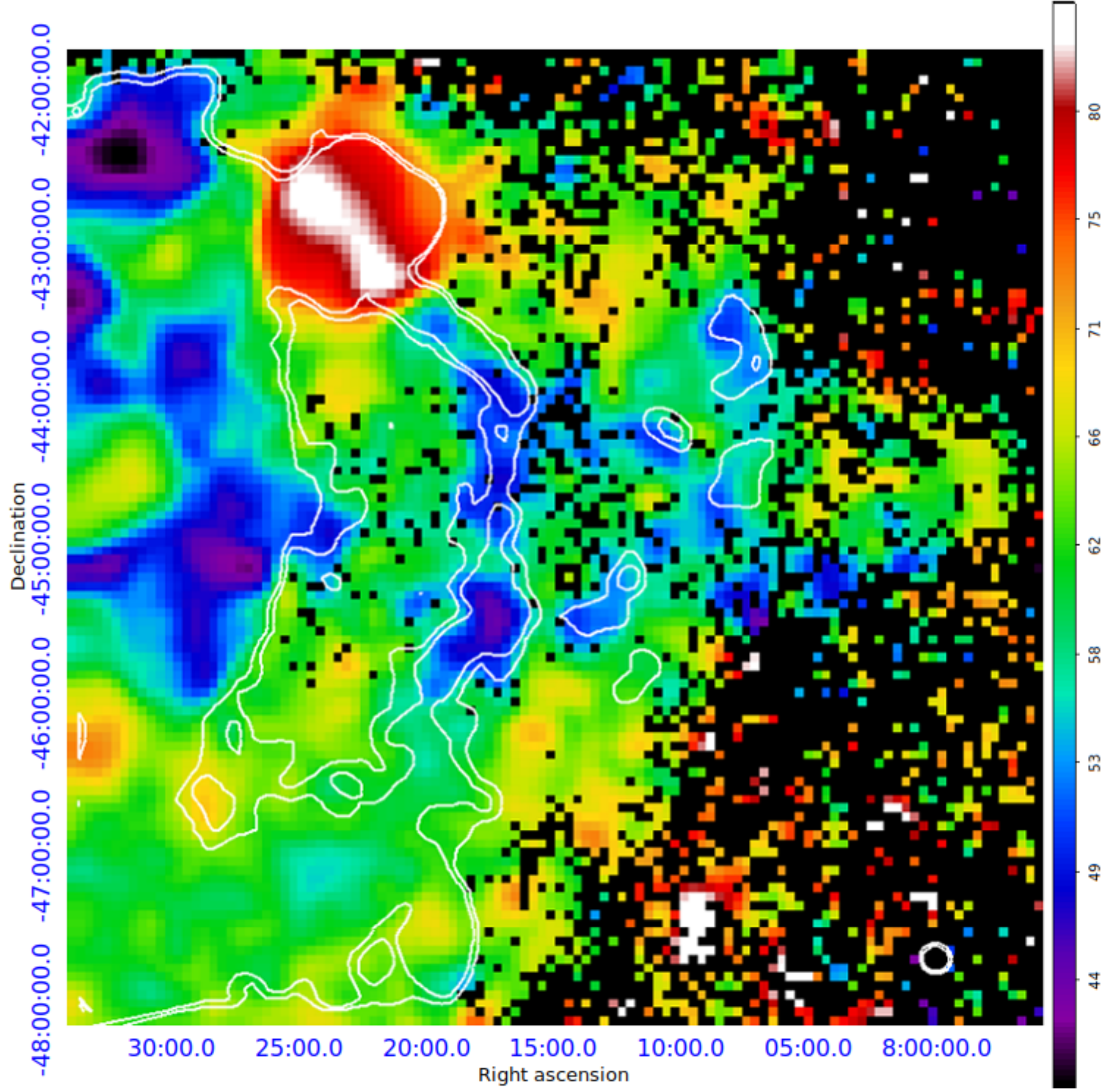}
     \caption{Median photon energy map of the western part of the Vela shell in the $0.3-1.3$ keV energy band with a bin size of $3.5'$ in linear scale.
     The scale is in units of 10 eV.
     White contours mark the photons'. count number levels between 4 and 5 counts in the $0.3-0.5$ keV energy band.
    }
     \label{fig:smomperos}
\end{figure}

In Fig. \ref{fig:smomperos} we show the $E_m$ map of the \emph{ROSAT} RASS observation in the $0.3-1.3$ keV energy band that we produced to further characterize the $E_m$ of the extended filament. 
The image shows that the X-ray emission of the extended filament is less energetic than that of the surrounding plasma. 
Also, the whole structure is coherent in $E_m$ on a very large spatial scale, comparable with the diameter of the shell. The relatively low $E_m$ of the extended filament ($450-550$ eV, indicated by the blueish structure in Fig. \ref{fig:smomperos}) presents only minor spatial variations that may not necessarily be caused by inhomogeneities in the plasma temperature. Variations in $E_m$ of the soft energy band considered, in fact, may be associated with different values of the absorbing column density, which is expected to vary on these large spatial scales (\citealt{2000A&A...362.1083L}).

\subsection{Spectra}
\begin{figure}
\centering
    \includegraphics[width=\columnwidth]{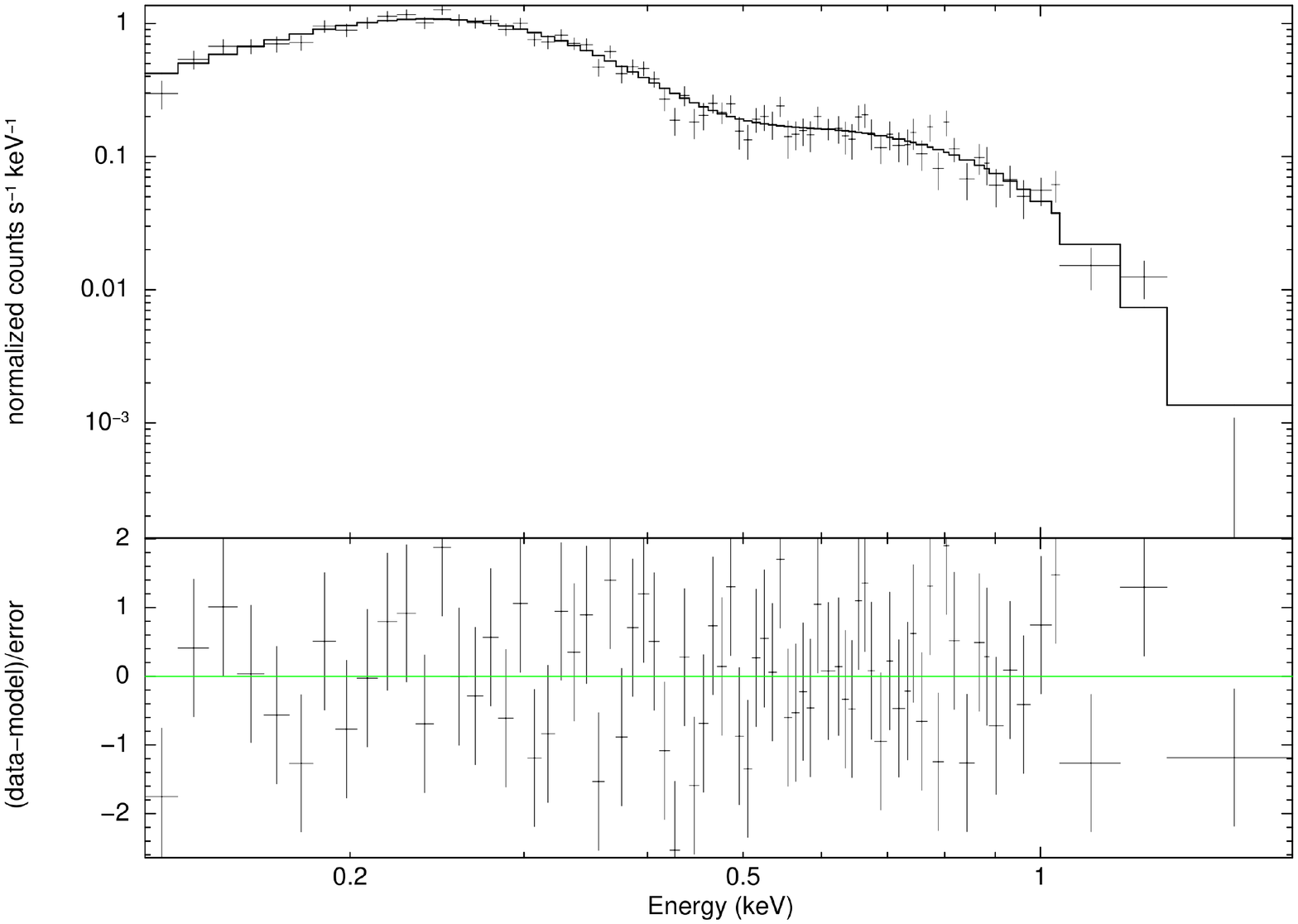}
    \caption{\emph{ROSAT} PSPC-C spectrum of region \emph{Ext\_fil} (black polygon in Fig. \ref{fig:ROSAThardsoft}, \emph{center panel}) with the best-fit CIE model  in the $0.3-1.8$ keV energy band. The bottom panel shows the residual between the data and the model.}
    \label{fig:extfilspec}
\end{figure}

The \emph{ROSAT} image analysis suggests that the narrow filament detected in the \emph{XMM-Newton} observation may be a portion of a much larger structure that clearly sticks out in Fig. \ref{fig:ROSAThardsoft}.
We extracted the spectrum from the region indicated by the black polygon, labeled \emph{Ext\_fil}, in Fig. \ref{fig:ROSAThardsoft}; we also extracted the background spectrum from the circular region indicated in red in Fig. \ref{fig:ROSAThardsoft}.
We fit the spectrum by adopting the best-fit model that we found for the \emph{XMM-Newton} filament, that is CIE thermal emission from an isothermal plasma (\texttt{vapec}) with O and Ne abundances free to vary, and including the effects of ISM absorption (\texttt{tbabs}).
Despite the poor \emph{ROSAT} PSPC energy resolution, we were able to derive the best-fit parameters  independently from the \emph{XMM-Newton} spectral analysis of the filament (see Tab. \ref{tab:bestfilROS}).
The PSPC spectrum of \emph{Ext\_fil} with the corresponding best-fit model and residual is shown in Fig. \ref{fig:extfilspec}.

\begin{table}
        \centering
        \caption{\emph{ROSAT} filament best-fit.}
        \label{tab:bestfilROS}
        \medskip

        \begin{tabular}{cc}
                \hline \hline
                parameter&Ext\_fil
                \\
                \hline
                \vspace{0.1cm}
        $N_\text{H}$ ($10^{22}$\hspace{0.1cm} cm$^{-2}$)&$0.029_{-0.004}^{+0.006}$
       \\
       \vspace{0.1cm}
       $kT$ (keV)&$0.177_{-0.016}^{+0.011}$ 
       \\
       \vspace{0.1cm}
       O/O$_{\astrosun}$&$0.22_{-0.08}^{+0.09}$
       \\
       \vspace{0.1cm}
       Ne/Ne$_{\astrosun}$&$<1.5$
       \\
       \vspace{0.1cm}
       $n^2l$ ($10^{18}$ cm$^{-5}$)&$0.017_{-0.005}^{+0.005}$
       \\
       \vspace{0.1cm}
       $\chi ^2/d.o.f$&93.15/94
       \\
       \hline

        \end{tabular}
         \begin{threeparttable}   
\begin{tablenotes}
\item The abundance of O was frozen to the best-fit parameter in order to reduce the uncertainty of the kT parameter. The abundance of O was determined by freezing Ne abundance.
        Errors are at a 90\% confidence level.
   \end{tablenotes}
   \end{threeparttable}
\end{table}

\begin{figure}
    \centering
    \includegraphics[width=\columnwidth]{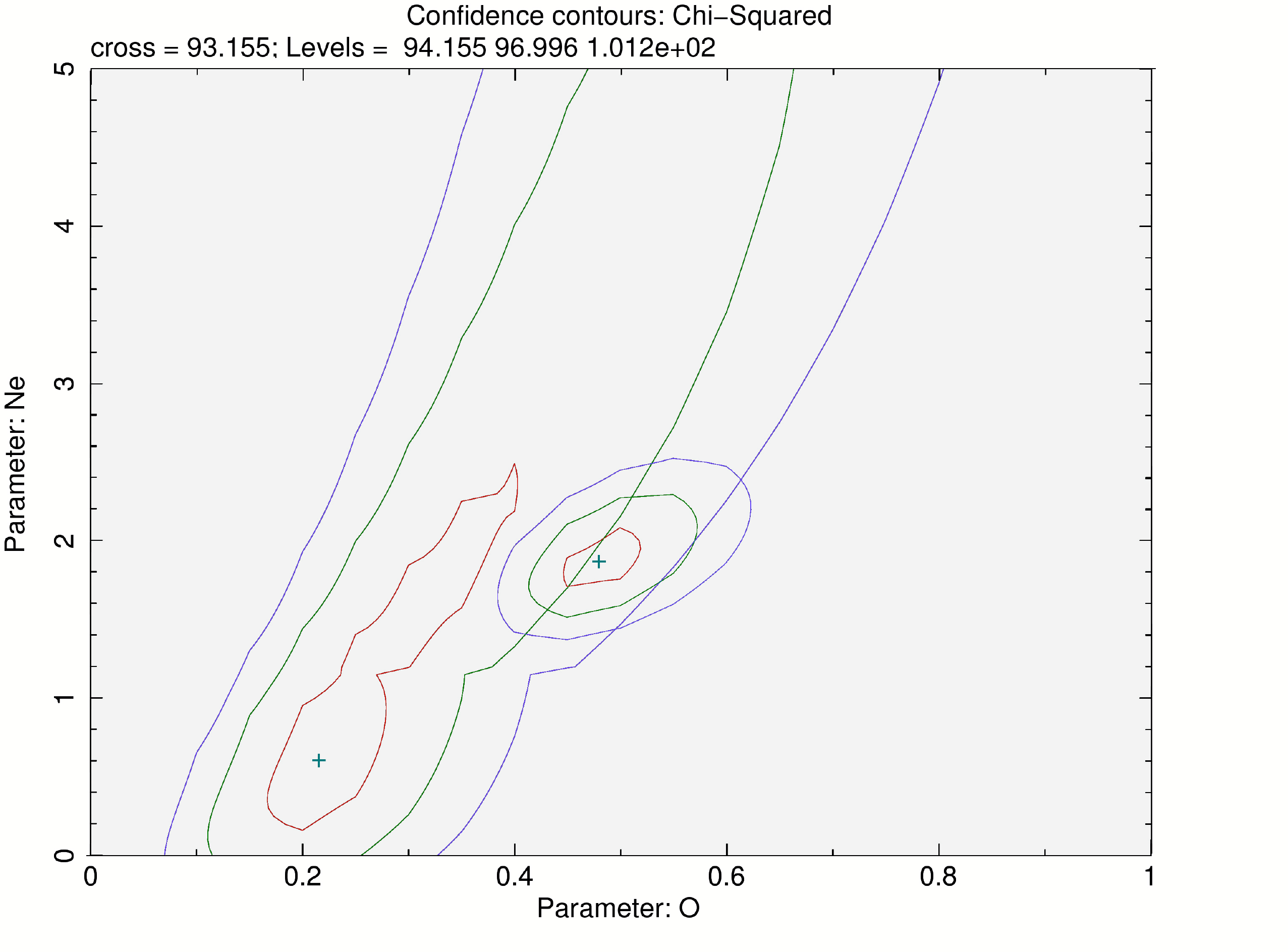}
    \caption{68\% (red), 95\% (green), and 99.5\% (blue) contour levels of the Ne abundance vs. O derived from the spectral analysis of the \emph{Ext\_fil} spectrum with the same contour levels of \emph{fil\_all} spectrum superimposed.}
    \label{fig:blendbanana}
\end{figure}

The best-fit temperature of the large filament is remarkably consistent with that obtained for the narrow filament in the \emph{XMM-Newton} observation ($kT_{fil\_all}=0.174_{-0.002}^{+0.002}$ $kT_{Ext\_fil}=0.177_{-0.016}^{+0.011}$, see also Tab. \ref{tab:filbest} and Tab. \ref{tab:bestfilROS} for comparison).
It is important to note that Ne and O abundances are not well constrained because of the low energy resolution of the PSPC-C \emph{ROSAT} spectra.
Fig. \ref{fig:blendbanana} shows the 68\%,  95\%, and 99.5\% confidence contour levels of the Ne abundance versus the O abundance derived from the \emph{ROSAT} spectrum of the large filament compared to the levels for the narrow filament in the \emph{XMM-Newton} data.
The plot shows that abundances are consistent within the 2$\sigma$ (green) confidence level, thus confirming the homogeneous chemical composition of the filament.
These results strongly indicate that the narrow filament is part of a coherent, bent, giant structure, running north-south behind the eastern rim of the Vela shell.


\section{Discussions}
\label{sect5}
In this paper, we have presented a detailed study of an \emph{XMM-Newton} observation of a bright X-ray emitting clump, namely knot K, located behind shrapnel G in the southwestern region of \vel\ (see Fig. \ref{fig:VELA}).
By analyzing  the \emph{XMM-Newton} observation, we found an X-ray emitting plasma structure (predominantly in the $0.3 - 0.5$ keV energy band, Fig. \ref{fig:rbimage}), which was remarkably elongated in the direction connecting shrapnel A and shrapnel G (see Fig. \ref{fig:mosaic}) (i.e., the only two  Si-rich shrapnel detected in the Vela SNR so far). Furthermore, the elongated knot points toward the explosion site of the \vel.
This structure shows a soft thermal spectrum with  nonsolar abundances, as shown in Fig. \ref{fig:knotspec}, and a temperature significantly lower than that of shrapnel A and G.
Although the elongated knot has a low temperature that hampers the detection of over-solar Si abundance, we found that the Ne to O abundance ratio is consistent with that of shrapnel A (\citealt{2006ApJ...642..917K}) and shrapnel G (\citealt{2017A&A...604L...5G}). 
Moreover, the Si abundance we found (though poorly constrained) is consistent with that observed in those ejecta fragments.
Enhanced Ne/O abundance ratios have also been observed in ISM clumps within the Vela SNR (\citealt{2005A&A...442..513M}, \citealt{2011PASJ...63S.827K}), though with slightly lower values than those presented in Table \ref{tab:knotbest}. It is then possible that the elongated structure is a shocked ISM cloudlet. However, its highly elongated morphology seems to suggest an association with shrapnel G. Moreover, its chemical composition is consistent with being the same as that of shrapnel A and shrapnel G.
In summary, our results show that a physical relationship between the jet-like elongated knot and the two Si-rich shrapnel is likely.

Assuming that the elongated knot has a cylindrical symmetry, we calculated the volume ($V$) of the X-ray emitting structure  through the relation $V = \pi \frac{D^2}{4} l$ where $l$ is the projected length of the elongated knot ($l\sim5\times10^{18}$ cm) and $D$ is its projected thickness ($D\sim1\times10^{18}$ cm), thus obtaining $V\sim4\times10^{54}$ cm$^3$.
We point out that from the X-ray image, we can only measure the projected size of the features in the plane of the sky, so this value should be considered as a lower limit. However, since the structure is close to the border of the shell, the actual value may not differ too much from the one reported here.
Through the best-fit value of the plasma emission measure and volume, we estimated a number density of $\overline{n} =1.20_{-0.07}^{+0.05}$ cm$^{-3}$ for the elongated knot and a total mass of $M = 1.00_{-0.04}^{+0.06}\times 10^{31}$ g ($\sim0.005M_{\astrosun}$), using as average atomic mass $\mu=2.1 \times10^{-24}$ g (value for solar abundances). However this value is an upper limit since we expect oversolar chemical abundances\footnote{The moderate spectral resolution of CCD spectrometers hampers the possibility of measuring absolute abundances and only reliable estimates of relative abundances can be obtained (see \citealt{2020A&A...638A.101G}).}.

Considering the projected distance between the explosion  center of \vel\ and the elongated knot and the age of the remnant ($\simeq$ 11000 yrs, \citealt{1993ApJS...88..529T}), we obtained v$_k\sim1.2\times10^8$ cm/s (1200 km/s) and a kinetic energy of $E_k = 7.2_{-0.4}^{+0.3}\times10^{46}$ erg.
However, if we take into account that the speed of the elongated knot decreases with time and accounts only for the projected velocity, this value of kinetic energy should be considered as a lower limit.
Values of mass and kinetic energy found for the elongated knot are very similar to those found for shrapnel G by \citet{2017A&A...604L...5G}.
By assuming that the elongated knot is part of a jet-counterjet-like structure with shrapnel A and G, its total mass and kinetic energy are $M = 0.018~ M_{\astrosun}$ and $E = 4.7\times10^{47}$ erg, respectively. 

\begin{figure}
    \centering
    \includegraphics[width=\columnwidth]{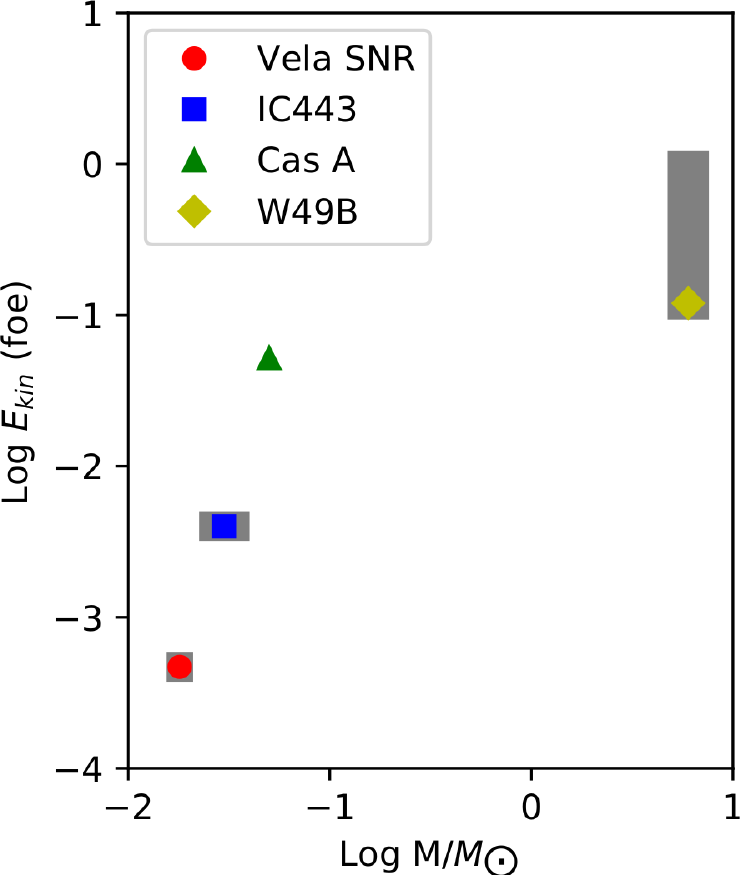}
    \caption{Kinetic energy vs. mass of the jet-like structure observed so far in core collapse SNRs. Kinetic energy is expressed in foe (1 foe = $10^{51}$ erg).
    Values are from this paper, \cite{2006ApJ...642..917K} and \cite{2017A&A...604L...5G} (\vel), \cite{2018A&A...615A.157G} (IC443), \cite{2016ApJ...822...22O} (Cas A), and \cite{2008AdSpR..41..390M} (W49B).
    }
    \label{fig:jetlike}
\end{figure}

Similar jet-like features have been observed in a handful of core-collapse SNRs and may be associated with anisotropies in the SN explosion (\citealt{2013MNRAS.430.2864M}, \citealt{2015MNRAS.453..166T}, \citealt{2016ApJ...822...22O}, \citealt{2018ApJ...855...82B}, \citealt{2020A&A...642A..67T}, ).
\cite{2008AdSpR..41..390M} found an indication of a jet-like structure for the galactic SNR W49B with mass $M=6M_{\astrosun}$ and $E=1.2\times10^{50}$ (see also \citealt{2013ApJ...777..145L}). However, it has been suggested that this collimated structure may not be intrinsic and may result from a (nearly) spherical explosion in a barrel-like circumstellar environment (\citealt{2008AdSpR..41..390M}, \citealt{2011MNRAS.415..244Z}, \citealt{2018A&A...615A.150Z}).
Also the Cas A SNR, deeply analyzed by \cite{2012ApJ...746..130H}, shows a jet of Si-rich plasma.
\citet{2006ApJ...644..260L} suggest that its origin lies in an explosive jet.
Detailed 3-D hydrodynamic simulations by \citet{2016ApJ...822...22O} proved that this jet can be explained as the result of velocity and density inhomogeneities in the ejecta profile of the exploding star. According to these simulations, the mass and kinetic energy of the jet-counterjet structure in Cas A are $M=0.05M_{\astrosun}$ and $E=5.2\times10^{49}$ erg, respectively.
Another jet-like feature in the ejecta has been recently discovered in the SNR IC 443 by \cite{2018A&A...615A.157G} who identified a jet-like structure with Mg-rich plasma in overionization with  mass and  kinetic energy  $M\sim0.03M_{\astrosun}$ and $E\sim3\times10^{48}$ erg, respectively.
A comparison between the \vel\ and the other jet-like structures observed in core-collapse SNRs shows a wide range of masses and energies, as shown in Fig. \ref{fig:jetlike}.
A simple linear regression gives $E=(4.7\pm1.7)\times10^{49}$ erg  $\cdot M/M_{\astrosun}$, but large residuals are present and the number of data points is too limited to get robust information.
We conclude that the morphology, the position, and the spectral analysis strongly suggest that the elongated knot is part of a knotty, collimated ejecta structure and/or an ejecta trailing wake left behind the supersonic motion of shrapnel G.

The $0.5-1.0$ keV count-rate image of the \emph{XMM-Newton} observation dominantly shows the emission of a narrow filamentary plasma structure (see Fig. \ref{fig:rbimage}).
The filament has a thermal spectrum (Fig. \ref{fig:filspec}) and its temperature is significantly higher than that of the elongated knot and uniform along its whole length (see Fig. \ref{fig:mpe03to10}).
The Ne to O ratio (Ne/O$=4.0_{-0.8}^{+0.6}$) is significantly higher than that of the elongated knot and, in general, than that of the ejecta fragments of Vela SNR. This indicates that the filamentary structure may have a different nature.

We demonstrated, thanks to the RASS data of \vel, that this X-ray emitting plasma structure is the southern edge of a more extended filament that runs from the northern rim of the shell to knot K (see Fig. \ref{fig:ROSAThardsoft}).
The extended filament is a giant X-ray emitting structure with a uniform temperature, whose physical and chemical properties are consistent with those measured in the \emph{XMM-Newton} filament (as shown in Tab. \ref{tab:filbest}, Tab. \ref{tab:bestfilROS}, and Fig. \ref{fig:blendbanana}).
The projected length of this extremely long structure is $l\sim6\times10^{19}$ cm.
The distance between the explosion site and the extended filament shows relatively small variations, ranging from $\sim4.5\times10^{19}$ cm to $\sim5.5\times10^{19}$ cm.

As shown with the \emph{ROSAT} results, this long filament is located well behind the border of the shell and its X-ray emission is softer than that of the forward shock (see Fig. \ref{fig:smomperos}). 
Soft emission may be the result of an interaction with a dense environment, given that the shock speed scales as the inverse of the square root of the particle density and the post-shock temperature increases as the square of the shock speed. 
This indication, together with the almost circular shape, centered at the position of the SN progenitor, suggests that the filament may be a dense stellar wind-blown relic of the progenitor star heated by the shock wave.

Figure \ref{fig:smomperos} also shows a high $E_m$ belt running through the SNR Puppis A, exactly at the same position as the large-scale filament we analyzed in this paper. \cite{2006A&A...454..543H} suggest that this belt may be due to intervening absorbing material from \vel.
\cite{2013A&A...555A...9D} present a comparison of the X-ray image of SNR Puppis A with the HI column distribution (\citealt{2003MNRAS.345..671R}) that shows a stripe with enhanced hydrogen column density ($N_\text{H}\sim10^{21}$ cm$^{-2}$) in coincidence with the high $E_m$ belt, suggesting that dense material might be responsible for absorbing soft X-ray photons.
Given the spatial coincidence of the large scale filament with the high absorption feature in Puppis A, it may be possible that the filament, together with the X-ray emitting plasma, also includes cooler and denser material, which may be responsible for such a large absorption.
This scenario surely requires further investigation of the lower energy emission from this feature.

In this scenario, the filament can be considered as the projection on the plane of the sky of a nearly spherical shell, likely a wall of a wind-blown bubble.
Given the poor spatial resolution of the \emph{ROSAT} telescope, the width of the shell can be estimated by the \emph{XMM-Newton} data only (i.e., only in the southern edge of the filament) and is $W\sim2\times 10^{17}$ cm. 
In the following, we assume that this is the width along the whole filament.
We estimate the volume of the X-ray emitting plasma within the photon extraction regions, which we approximated as the volume intercepted by the extraction region on two spherical shells (having width $W$) and different radii, $ r_1=5.0\times10^{19}$ cm and $r_2=4.3\times10^{19}$ cm. This can be done by adopting the procedure described in \citet{2012A&A...546A..66M} (see their Appendix A). 

We obtain a density $n_{fil}\sim0.182_{-0.015}^{+0.020}$ cm$^{-3}$ for the whole filament (region \emph{Ext\_fil} in Fig. \ref{fig:ROSAThardsoft}, analyzed with \emph{ROSAT}) and $n_{fil}\sim 0.40\pm0.02$ cm$^{-3}$ for the southern edge of the filament (region \emph {fil\_all} in Fig. \ref{fig:specreg}, analyzed with \emph{XMM-Newton}). This may indicate small (within a factor of $\sim2$) spatial variations in the shell density and/or variations in its width (the assumption of a uniform width may not be strictly valid and this may affect the density estimate).  
By considering an average density $n_{fil}=0.2$ cm$^{-3}$, we also estimated the mass of the spherical shell blown by the wind with external radius $R_{ext}=4.49\times10^{19}$ cm and internal radius $R_{int}=4.47\times10^{19}$ cm, thus finding $M_{shell}=2.5_{-0.2}^{+0.3}\times10^{33}$ g ($1.26_{-0.10}^{+0.15}$ $M_{\astrosun}$).

Considering the projected average distance between the explosion center of \vel\ and the forward shock in the western region, we obtained a velocity of $v\sim2\times10^8$ cm/s. Using this velocity and considering an average distance from the explosion center ($r_{shell}\sim5\times10^{19}$ cm), we found that the shock interacted with the filament structure approximately 8 kyr after the explosion.
This value is an upper limit value. 
Indeed, the filamentary structure was probably dragged by the interaction with the forward shock.
It is also possible to get an estimate of the age of this putative wind-blown structure. By using the $r_{shell}$ value and the characteristic speed of stellar winds in a red supergiant ($\sim10$ km/s), we determined that the progenitor star started to blow this wind (i.e., entered in its red supergiant phase) $\sim1.6$ Myr ago. The onset of the red supergiant phase typically precedes the explosion by $0.5-2$ Myr; therefore, our time estimate is quite reasonable and strongly supports the interpretation of the filament as a wind-blown structure produced by the progenitor star. 

On the other hand, the high Ne/O ratio suggests peculiar abundances for this putative wind-blown shell. A possible explanation for a high Ne/O ratio can be O depletion. In layers where the CNO cycle is active, the O abundance strongly decreases (while Ne abundance stays steady). Therefore, large Ne/O abundance ratios are expected in the inner part of the H layer of a massive star. To blow a wind with enhanced Ne/O abundances, it is then necessary for the star to lose its H envelope, as occurs in Wolf-Rayet stars (WNL-WNE). 
A Wolf-Rayet star can blow a fast wind that can reach, and impact, the previously ejected material, thus forming a shell with enhanced Ne/O abundances, similar to what we observed (\citealt{2013ApJ...764...21C}, \citealt{2018ApJS..237...13L}). In order to test this scenario, it would be necessary to measure the N and C abundances in the wind-blown shell. Unfortunately, we cannot obtain significant constraints on these abundances with our data and further investigations are necessary. The detection of a wind-blown relic, in any case, can convey important information on the progenitor star of Vela SNR.

\begin{acknowledgements}
The authors are grateful to the referee for very constructive comments and inspiring suggestions.
The authors would like to thank A. Chieffi and M. Limongi for the very helpful suggestions regarding the metallicity in stellar winds.
MM acknowledges partial support from the Italian Ministry of Research through the "Fondo per il finanziamento delle attività base di ricerca" (FFABR 2017).
SO, MM, FB and EG acknowledge financial contribution from the INAF mainstream program.
FG and JAC acknowledge support by PIP 0102 (CONICET). JAC is CONICET researcher. This work received financial support from PICT-2017-2865 (ANPCyT). JAC was also supported by grant PID2019-105510GB-C32/AEI/10.13039/501100011033 from the Agencia Estatal de Investigaci\'on of the Spanish Ministerio de Ciencia, Innovaci\'on y Universidades, and by Consejer\'{\i}a de Econom\'{\i}a, Innovaci\'on, Ciencia y Empleo of  Junta de Andaluc\'{\i}a as research group FQM-322, as well as FEDER funds.
FG acknowledges the research programme Athena with project number 184.034.002, which is (partly) financed by the Dutch Research Council (NWO).
\end{acknowledgements}

\bibliographystyle{aa}
\bibliography{biblio_master}

\end{document}